\documentclass[a4paper,11pt]{article}
\pdfoutput=1
\IfFileExists{jheppub2.sty}{\usepackage{jheppub2}}{%
  \IfFileExists{jheppub.sty}{\usepackage{jheppub}}
  {\usepackage{jheppub-fallback}}}
\makeatletter
\@ifpackageloaded{natbib}{}{\usepackage[numbers,sort&compress]{natbib}}
\makeatother
\providecommand{\doi}[1]{\href{https://doi.org/#1}{doi:\nolinkurl{#1}}}

\usepackage{amsmath,amssymb,mathtools,bm}
\usepackage{booktabs,array,microtype,enumitem}
\usepackage[section]{placeins}

\usepackage[capitalise,nameinlink]{cleveref}

\numberwithin{equation}{section}
\setlist{itemsep=2pt,topsep=3pt}
\setlist[itemize]{label=$\bullet$}

\newcommand{\dd}{\mathrm{d}}
\newcommand{\e}{\mathrm{e}}
\newcommand{\sech}{\operatorname{sech}}
\newcommand{\Ib}{\bar I}
\newcommand{\xcl}{x_{\rm cl}}
\newcommand{\Dh}{D_{\rm half}}
\newcommand{\Dvac}{D_{\rm vac}}
\newcommand{\kq}{\kappa_q}

\title{Algebraic instantons and  non-Gaussian vacuum in quantum mechanics}
\author[a]{Rahim Aziz,}
\author[b]{Hao Geng,}
\author[a]{Thomas Sch\"afer,}
\author[a]{and Mithat \"Unsal}
\affiliation[a]{Department of Physics, North Carolina State
University, Raleigh, NC 27695, USA.}
\affiliation[b]{Gravity, Spacetime, and Particle Physics (GRASP) 
Initiative, Harvard University, 17 Oxford St., Cambridge, MA,
02138, USA.}
\emailAdd{rabalsar@ncsu.edu}
\emailAdd{haogeng@fas.harvard.edu}
\emailAdd{tmschaef@ncsu.edu}
\emailAdd{unsal.mithat@gmail.com}

\abstract{The standard instanton calculus in quantum mechanics
is based on harmonic minima in the potential. When the perturbative
vacuum is anharmonic, $V\simeq\lambda x^{2n}$ with $n\ge2$, the
standard construction fails. We study this problem for a particle
on a circle and subject to a potential $V(x)=16\lambda\sin^4(x/2)$.
The instanton and its fluctuation operator are known in closed form,
but they differ qualitatively from the familiar ones. Because $V''=0$
at the minimum, the classical vacuum is gapless, the instanton
decays algebraically (an \emph{algebraic instanton}), the zero mode
is located at the threshold of a continuous spectrum, and the
one-loop determinant is infrared divergent. Quantum mechanics
generates a non-perturbative gap $E_q\propto\hbar^{4/3}\lambda^{1/3}$,
and the instanton is embedded in a vacuum that is dynamically gapped,
similar to what happens in Yang--Mills theory. We show that the
one-instanton amplitude can be computed by matching the instanton
trajectory to the exact vacuum part of the path integral. The
resulting prefactor scales as $\hbar^{4/3}$, and this expression
agrees with the numerical spectrum to a very high accuracy. The
classical interaction between instantons decays as $T^{-3}$, whereas
the correct result is an exponential decay controlled by the
dynamically generated gap, beyond the quantum time scale. We provide
an alternative derivation of this result based on a sum over
intermediate states. }


\begin{document}
\maketitle
\flushbottom

\section{Introduction}
\label{sec:overview}

In this paper we introduce and study instantons in a 
quantum mechanical set-up that is generic in field theory:
\begin{itemize}
\item the instanton is embedded in a vacuum that is gapless 
classically, but a gap is formed dynamically. This is similar 
to the instanton in Yang--Mills theory on $\mathbb R^4$, and 
to sigma models on $\mathbb R^2$.
\end{itemize}
Interestingly, this class of instantons has not been studied in 
depth in quantum mechanics (QM), although it is probably as generic 
as the standard case, and, because of the nature of the vacuum 
it sits in, perhaps closer to the instantons of field theory. 
The prototypical QM examples we have in mind are the \emph{quartic} 
double well and the quartic periodic potential,
\begin{align}
V(x)&=\lambda\,x^4(1-x)^4, && x\in\mathbb R, 
\label{eq:dw}\\
V(x)&=16\lambda\sin^4(x/2), && x\equiv x+2\pi k,\quad k=1,2,\dots 
\label{eq:periodic}
\end{align}
The instantons in these two systems are very similar. We will
work with the second, with $k=1$, because it is the simplest, 
and return to the first in \cref{sec:dw}.

\paragraph{The standard lore (harmonic minima).} 
The instanton calculus in quantum mechanics, as described 
in standard textbooks \cite{Polyakov,Coleman,ABC,ZJ,Rajaraman}, 
is built on the double well and on periodic potentials whose minima
are harmonic \cite{GildenerPatrascioiu}, and it is well understood. 
All-orders perturbative and non-perturbative results around the 
perturbative and instanton saddles \cite{ZJJ2004,ZJJ2004b,
Alvarez2004,DunneUnsal,DunneUnsal2,DunneUnsalMathieu,BDU,MNS,
FKMNS,MarinoBook}, the role of complex saddles and Lefschetz 
thimbles \cite{CDU,HTA,CuriousIncident,BDSSU,KSTU}, and the 
exact-WKB analysis \cite{KawaiTakei,IwakiNakanishi,GGM,SKMU2020,
KMSU,BRS} have been worked out in detail. The double well itself 
continues to be revisited \cite{DersySchwartz,Shapiro}. For the 
points we make in this paper most of this machinery is not needed.
We only remind the reader that the one-instanton contribution to 
the tunneling amplitude is
\begin{equation}
t=\hbar K\,\e^{-S_0/\hbar},\qquad 
K=\sqrt{\frac{S_0}{2\pi\hbar}}
\left[\frac{\det' M}
      {\det\!\big(-\partial_\tau^2+\omega^2\big)}\right]^{-1/2},
\label{eq:coleman}
\end{equation}
where $M$ is the fluctuation operator around the instanton, the 
prime indicates that the zero mode is removed, and the denominator 
is the fluctuation operator around the \emph{Gaussian} vacuum.
Everything in \eqref{eq:coleman} is finite because the vacuum has a 
gap $\omega$: the instanton approaches the vacuum exponentially, 
$\dot x\sim A\,\e^{-\omega|\tau|}$, the zero mode is an isolated
eigenvalue, the ratio of determinants can be computed by the 
Gelfand--Yaglom method \cite{GY}, and the result agrees with 
the WKB evaluation of the splitting \cite{LL,Garg}. The 
multi-instanton sector rests on the same gap: well-separated 
instantons can be superposed because, although the instanton 
equations are \emph{non-linear}, outside their cores the deviation
from the vacuum obeys a \emph{linear} equation, and the resulting 
exponentially small interactions are what makes the dilute gas 
approximation and its multi-instanton refinements \cite{DunneUnsal,
DunneUnsal2,ArgyresUnsal,DunneUnsalCPN,BDSSU} work.

\paragraph{Anharmonic minima change almost everything.} For 
$V=16\lambda\sin^4(x/2)$ the minimum is quartic, $V\simeq
\lambda x^4$, so $\omega=0$. At the classical level nothing 
looks wrong: the solution of the instanton equation is known 
in closed form\footnote{It is known in $(1+1)$-dimensional 
field theory, as the kink of the $\cos^4\phi$ model 
\cite{KhareSaxena2022}; we are not aware of a treatment of 
it as an instanton.}, the action is finite, and the fluctuation 
operator is exactly solvable. The problem is related to the 
classical vacuum. With
\begin{equation}
V''\big|_{x_{\min}}=0
\end{equation}
the classical vacuum is gapless, and every step of the construction 
above that relied on the gap fails: the instanton has power-law tails, 
$x\sim1/|\tau|$ (an \emph{algebraic instanton}), the zero mode is a 
bound state sitting exactly at the threshold of a gapless continuum, 
and the determinant ratio in \eqref{eq:coleman} is infrared divergent, 
growing as $L^5$ on an interval of size $L$. Two instantons cannot be 
superposed. Indeed, the sum of two instantons is not an approximate 
solution anywhere, unlike the textbook examples with harmonic minima, 
and the action of the superposition has the wrong sign and the wrong
power of the separation in both the instanton--anti-instanton 
and the instanton--instanton channel. 

The fact that the naive dilute gas fails for non-harmonic minima was
noted long ago~\cite{Patrascioiu,MottolaRouet}. We find that the
instanton calculus is not lost, but the part of the path integral
where the instanton tail merges with the vacuum must be evaluated
exactly. In path-integral language, the vacuum segments
$\langle x'|\e^{-H\tau}|0\rangle$ are not Gaussian; at long times they
reduce to $\e^{-E_0\tau}\psi_{\rm vac}(x')$, so the exact ground-state
wavefunction of the quartic oscillator enters as the boundary factor
to which the instanton is joined. With this input the instanton gives
the correct amplitude, and the dilute gas survives because the gap is
generated dynamically.

None of this is special to the quartic case. For a potential near
the minimum of the form $V\simeq\lambda x^{2n}$ the tail of the 
instanton obeys $\ddot\eta\propto\eta^{2n-1}$, which is linear 
only for $n=1$. For every $n\ge2$ the tails are power laws, $\eta\propto|\tau|^{-1/(n-1)}$; we call
such solutions \emph{algebraic instantons}, by analogy with the algebraic
solitons of nonlinear waves. The classical vacuum is gapless, and  superposition fails. In the family of $x^{2n}$ minima the 
failure of superposition is the rule, and the harmonic minimum, 
where it works, is the exception\footnote{For $V\propto
\sin^{2n}(x/2)$ the instanton is solvable in parametric form: 
with $u=x/2=2\arctan\e^{y}$, the equation $\dd u/\dd s=\sin^nu$ 
gives $s=\int_0^y\cosh^{n-1}y'\,\dd y'$, and the tails are 
$x\propto s^{-1/(n-1)}$. The separatrix argument of 
\cref{sec:separatrix} gives a $T^{-(n+1)/(n-1)}$ interaction 
with $|S_{II}/S_{I\Ib}|=[\sin\frac{\pi}{2n}]^{-2n/(n-1)}$, which 
is $4$ for $n=2$ and tends to $1$ as $n\to1$. The family is 
nevertheless less useful than it looks, because the instanton 
picture melts quickly as $n$ grows. The width of the ground 
state of the well, $\ell_n=G^{-1/(2n+2)}\sim S_0^{-1/(n+1)}$, 
grows with $n$, so the window $\ell_n\ll x\ll1$ in which the 
instanton tail is classical while the vacuum is already 
semiclassical closes: at $S_0=100$, $\ell_2\approx0.2$, 
$\ell_5\approx0.5$ and $\ell_{10}\approx0.7$. For $n$ of order 
a few the zero-point motion fills the well, which becomes a box
with thin walls, and there is no instanton tail to match to anything. 
At fixed barrier height $V_{\max}$ the action itself decreases, 
$S_0/\hbar\simeq4\sqrt{\pi\hat G/n}$ with $\hat G=IV_{\max}/
\hbar^2$, and the instanton picture is lost altogether for
$n\gtrsim16\pi\hat G$.}.
The same phenomenon has been observed for kinks with power-law 
tails in scalar field theory, where the naive superposition gives 
the wrong sign of the force \cite{Lohe,Manton,Christov,
KhareSaxena2019,GaniLenskyLizunova}. The quartic case is the 
simplest, and the one where both the failure of the textbook 
calculation and its cure can be exhibited cleanly. In this
work we concentrate on it.

\begin{figure}[t]
\centering
\includegraphics[width=\textwidth]{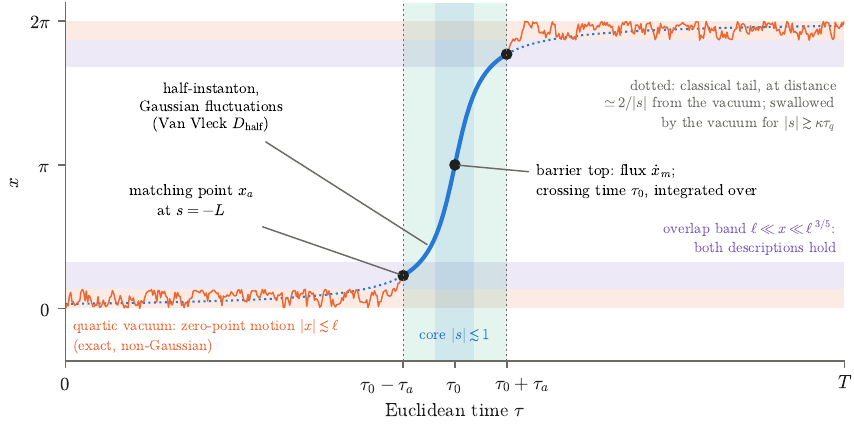}
\caption{The big picture for an instanton in a gapless potential: 
a schematic tunneling history ($\ell$ exaggerated). Near the
top of the barrier the path is the classical instanton, with Gaussian 
fluctuations. Near the minimum the power-law tail (dotted) 
disappears in the zero-point motion of the quartic well (orange 
strips, $|x|\lesssim\ell$), and the path is described by the exact, 
non-Gaussian ground-state wavefunction. The two descriptions overlap 
in an intermediate band (purple strips, $\ell\ll x\ll\ell^{3/5}$), 
where the wavefunction decays as $e^{-W(x)}$, with $W(x)$ the action 
of the instanton tail. We cut the history at $x_a$ and $2\pi-x_a$ in 
this band, i.e.\ at the times $\tau_0\mp\tau_a$ (green window): 
between the cuts are two half-instantons glued at the barrier top, 
with the zero mode traded for the crossing time $\tau_0$ weighted by 
the flux $\dot x_m$; outside, the ground-state wavefunction. The 
result is independent of $x_a$. The corresponding factorization of 
the path integral is shown in \cref{fig:cut}.}
\label{fig:picture}
\end{figure}

What replaces the gap, where does the classical instanton stop 
being a good description of the path, and what is the correct 
one-instanton amplitude? In fact, the answers to all of these 
questions are governed by a single quantum scale. Quantum 
mechanics generates a gap inside the well, $E_q\propto
\hbar^{4/3}\lambda^{1/3}$, with a corresponding width $\ell$ 
and time duration $\tau_q$. These scales are non-perturbative 
in the ordinary sense: there is no small parameter in the quartic
oscillator, since the coupling can be scaled out and only 
sets the overall units \cite{Simon,TurbinerDelValle}. This
is the same situation that motivates the $1/N$ expansion in 
atomic physics and in QCD \cite{WittenPT}. Once this scale 
has been determined for the parameter-free quartic oscillator,
everything in the instanton calculation that was infrared divergent 
is cut off at $\tau_q$ (\cref{fig:picture}): the tail of the 
instanton disappears into the zero-point motion of the vacuum, 
the power-law interaction between instantons turns into the 
exchange of a gapped excitation, and the Gaussian vacuum 
determinant is replaced by the vacuum wavefunction.

\paragraph{Main results.} We work in units $\hbar=I=1$, where 
$I$ is the moment of inertia for the classical rotor, so that 
the model has one coupling $G=I\lambda/\hbar^2$, the quartic 
coefficient at the minimum, and the instanton action is
$S_0=4\pi\sqrt{2G}$. The main observations are the following:
\begin{enumerate}[leftmargin=18pt,itemsep=3pt]

\item \emph{The problem with the naive one-loop calculation
is in the infrared, and it is related to the vacuum of the 
quartic oscillator, not the instanton.} The instanton, its 
fluctuation operator and its zero mode are all exactly solvable. 
The zero mode is a normalizable bound state at the threshold of 
a gapless continuum, and the zero-mode-removed determinant grows 
as a power of the size of the Euclidean time window. The divergence 
comes entirely from normalizing to a free, massless reference, 
that is, from treating the vacuum as Gaussian.

\item \emph{The vacuum of the quartic oscillator.} It has no 
small parameter, but it has definite scales: a gap $E_q\propto
\hbar^{4/3}\lambda^{1/3}$, a width $\ell$ and a time $\tau_q$, 
and its ground-state energy and tail constant are universal 
numbers. Corrections to it form a series in $\hbar^{2/3}$ rather 
than $\hbar$.

\item \emph{The vacuum part of the path integral must be evaluated 
exactly.} We divide the path at matching points where the asymptotic
quartic vacuum wavefunction agrees with the under-barrier WKB 
solution. Between these points, we use two semiclassical 
half-instantons with finite Dirichlet determinants and take 
the barrier-crossing time as the collective coordinate. Outside 
them, we evaluate the path integral using the vacuum wavefunction. 
The dependence on the matching points cancels. The resulting 
tunneling amplitude agrees with the Herring formula which is 
expressed using the single-well wavefunction at the barrier top 
and, after including the first correction, with the numerical 
spectrum at the percent level. Its prefactor scales as 
$\hbar^{4/3}$, without the usual factor $\sqrt{S_0/\hbar}$: it 
is determined by matching the semiclassical barrier region to the 
non-Gaussian quantum vacuum.

\item \emph{The determinant formula is not wrong, the vacuum factor 
is.} Replacing the Gaussian vacuum determinant by an effective
determinant built from the vacuum wavefunction restores it exactly. 
In general the vacuum enters an instanton amplitude through its
wavefunction at the matching point, only for a Gaussian vacuum does 
this reduce to a normalizing determinant in the denominator.

\item \emph{Instantons interact by a power law, until the vacuum 
screens them.} Superposition fails, but the exact classical 
interaction follows from the separatrix of the Euclidean motion: 
it decays as $T^{-3}$ in both $II$ and $I\bar{I}$ channels, with universal coefficients, 
the $II$ repulsion being four times the $I\Ib$ attraction. Beyond 
the quantum time $\tau_q$ it is replaced by the exchange of the 
gapped excitation of the well, so the dilute gas survives. The 
same interaction can be read as a sum over the states of the well.

\item \emph{The $\theta$-dependence interpolates smoothly} between the 
free rotor and the dilute instanton gas, with the crossover at
$S_0\approx3.5$.

\end{enumerate}

We note that several structural features of this system also appear 
in Yang--Mills theory on $\mathbb R^4$: instantons have power-law
tails, the classically gapless vacuum acquires a dynamically 
generated gap, and classical instanton interactions decay as a 
power law. In \cref{sec:YM}, we discuss these similarities and 
the scope and limitations of the analogy.

\section{Set-up, units and scales}\label{sec:setup}

Consider a particle on a circle parameterized by an angle 
$x\sim x+2\pi$. The classical moment of inertia is $I$, and
the Hamiltonian is
\begin{equation}
H=\frac{p^2}{2I}+16\lambda\sin^4\frac x2 .
\end{equation}
The potential has a single minimum on the circle, at $x=0$, 
where $V\simeq\lambda x^4$, and a barrier of height $16\lambda$ 
at $x=\pi$. The factor $16$ is chosen so that $\lambda$ is the 
quartic coefficient at the minimum. An instanton winds once 
around the circle, from $x=0$ to $x=2\pi$.

 Measuring energies in units of the rotor energy $\hbar^2/I$ and
Euclidean time in units of $I/\hbar$ gives
\begin{equation}
H=-\frac12\partial_x^2+16G\sin^4\frac x2,\qquad 
G\equiv\frac{I\lambda}{\hbar^2},
\label{eq:Hdimless}
\end{equation}
and the Euclidean action (in units of $\hbar$)
\begin{equation}
S_E=\int\dd\tau\,\Big[\tfrac12\dot x^2+16G\sin^4\tfrac x2\Big].
\end{equation}
With $x=2u$, $s=\kappa\tau$ and $\kappa=\sqrt{8G}$ the coupling 
factors out,
\begin{equation}
S_E=\frac{1}{\hbar_{\rm eff}}\int\dd s\,\big[u'^2+\sin^4u\big],
\qquad \hbar_{\rm eff}=\frac{1}{4\sqrt{2G}}=\frac{\pi}{S_0} .
\label{eq:heff}
\end{equation}
All classical quantities, measured in $s$, are therefore 
independent of $G$. The parameter $G$ enters only through the 
loop parameter $\hbar_{\rm eff}$, and through the quantum dynamics 
of the vacuum discussed in \cref{sec:vacuum}. From now on we set 
$\hbar=I=1$ unless stated otherwise. Physical units are restored by 
$G\to I\lambda/\hbar^2$, energies $E\to E\times\hbar^2/I$, 
and times $t\to t\times I/\hbar$.

Near the minimum $V\simeq G\,x^4$. The ground state of 
$-\tfrac12\partial_x^2+Gx^4$ has width, energy and time scale
\begin{equation}
\ell=G^{-1/6},\qquad E_q=\frac{1}{\ell^2}=G^{1/3},\qquad 
\tau_q=\frac{1}{E_q}=G^{-1/3}.
\label{eq:quantumscales}
\end{equation}
\Cref{tab:scales} collects these results together with the 
classical scales. In this region, the small parameter that controls 
corrections to the instanton picture is not $1/S_0$ but
\begin{equation}
\ell^2=G^{-1/3}=(32\pi^2)^{1/3}
   \Big(\frac{\hbar}{S_0}\Big)^{2/3}
 \simeq6.81\,\Big(\frac{\hbar}{S_0}\Big)^{2/3}.
\label{eq:ell2}
\end{equation}
Fractional powers of $\hbar$ are the first signature of the non-Gaussian
vacuum. The origin of these quantum scales can be understood from the
competition between kinetic energy and the quartic potential near
the minimum. Quantum fluctuations prevent the particle from being
localized exactly at the classical minimum, while the quartic
potential confines its motion. The balance between these two effects
determines the characteristic width of the vacuum wavefunction,
which in turn sets the energy and time scales. Unlike the harmonic
case, where the energy gap is determined by the classical curvature,
the quartic vacuum develops a finite gap entirely through quantum dynamics.

The hierarchy of time scales reveals the different physical regimes of 
the theory. The shortest scale, $\tau_I$, characterizes the classical 
instanton size (the counterpart of $\omega^{-1}$ for standard instantons 
with harmonic minima),  while $\tau_q$ marks the onset of quantum vacuum
effects. Between these scales, instantons interact through their classical 
algebraic tails. As we discuss in \cref{sec:interactions,sec:vacuum}, the
dynamically generated quantum gap screens these interactions at larger 
separations, replacing their power-law decay with exponential suppression.
Since $\overline{T} \gg \tau_q$, the dilute instanton gas
picture remains valid despite the algebraic classical tails.

\begin{table}[t]
\centering
\small
\begin{tabular}{@{}ll@{}}
\toprule
quantity & value ($\hbar=I=1$)\\
\midrule
barrier height $V_{\max}$ & $16G$\\
instanton action $S_0$ & $4\pi\sqrt{2G}$\\
instanton duration $\tau_I$ & $1/\kappa=1/\sqrt{8G}$\\
quantum width of the vacuum $\ell$ & $G^{-1/6}$\\
quantum energy $E_q$ & $G^{1/3}$\\
vacuum time $\tau_q$ & $G^{-1/3}$\\
ratio $\tau_q/\tau_I$ & $2\sqrt2\,G^{1/6}$\\
gap of the well & $(e_1-e_0)\,E_q=1.726\,G^{1/3}$\\
mean separation of instantons $\bar T$ 
        & $\simeq\tau_q\,\e^{S_0}/1.5$\\
expansion parameter around the instanton 
         & $\ell^2=G^{-1/3}\propto(\hbar/S_0)^{2/3}$\\
\bottomrule
\end{tabular}
\caption{Scales of the problem. The quantum scales $\ell,E_q,\tau_q$ 
are those of the pure quartic oscillator. The parameters $e_0,e_1$
are defined in \eqref{eq:quarticconstants}. The hierarchy 
$\tau_I\ll\tau_q\ll\bar T$ organizes the analysis.}
\label{tab:scales}
\end{table}

\section{The instanton}
\label{sec:instanton}

\begin{figure}[t]
\centering
\includegraphics[width=\textwidth]{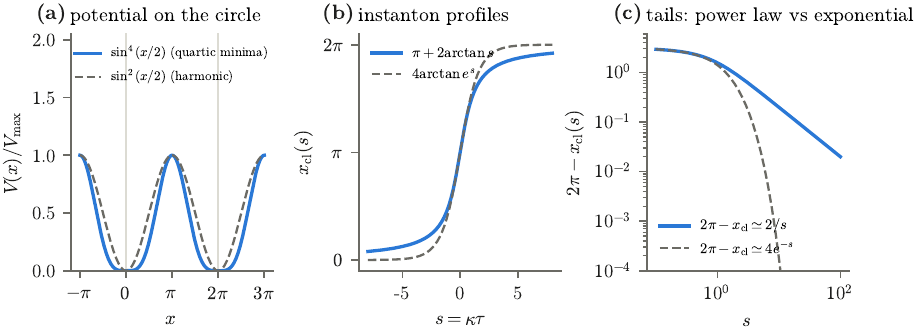}
\caption{(a) The quartic-minimum potential $\sin^4(x/2)$ 
compared to the harmonic $\sin^2(x/2)$. (b) The instanton 
\eqref{eq:instanton} and the sine-Gordon kink, both in units 
where the midpoint velocity is $2$. (c) The tails of the 
solution: $2/s$ versus $4\e^{-s}$.}
\label{fig:instanton}
\end{figure}

In Euclidean time the particle moves in the inverted potential 
$-V$. The instanton is the zero-energy trajectory from the 
hilltop at $x=0$ to the hilltop at $x=2\pi$:
\begin{equation}
\tfrac12\dot x^2=V(x)
\quad\Longrightarrow\quad 
\dot x=2\kappa\sin^2\frac x2,\qquad 
\kappa=\sqrt{8G}.
\end{equation}
Separating variables, $\int\dd x/(2\sin^2\frac x2)=-\cot\frac x2
=\kappa(\tau-\tau_0)$, so
\begin{equation}
\xcl(s)=\pi+2\arctan s,\qquad s=\kappa(\tau-\tau_0).
\label{eq:instanton}
\end{equation}
Indeed $\dd\xcl/\dd s=2/(1+s^2)=2\sin^2(\xcl/2)$. Its action is
\begin{equation}
S_0 = \int\dd\tau\,\dot x^2
    = \int_0^{2\pi}\sqrt{2V}\,\dd x
    = 2\kappa\int_0^{2\pi}\sin^2\frac x2\,\dd x
    = 2\pi\kappa=4\pi\sqrt{2G}.
\label{eq:S0}
\end{equation}
Half of the winding, from $\pi/2$ to $3\pi/2$, happens for 
$|s|<1$. The instanton has duration $\tau_I=1/\kappa$. Far 
from the core, however,
\begin{equation}
\xcl \simeq \frac{2}{|s|}\quad (s\to-\infty),\qquad 
2\pi-\xcl \simeq \frac2s\quad(s\to+\infty),
\label{eq:tails}
\end{equation}
a \emph{power law}. The reason is local: near the top of the 
inverted potential $\dot x\simeq\kq x^2$ with $\kq\equiv\sqrt{2G}
=\kappa/2$, with solution $x=1/(\kq|\tau|)$. For harmonic minima 
the same argument gives $\dot x\simeq\omega x$ and exponential 
tails. \Cref{fig:instanton} compares the situation with the 
sine-Gordon kink of a harmonic potential $\propto\sin^2(x/2)$, 
$\xcl=4\arctan\e^{s}$, with tails given by $4\e^{-|s|}$.

\section{Fluctuations, the zero mode and the determinant}
\label{sec:det}

\begin{figure}[t]
\centering
\includegraphics[width=\textwidth]{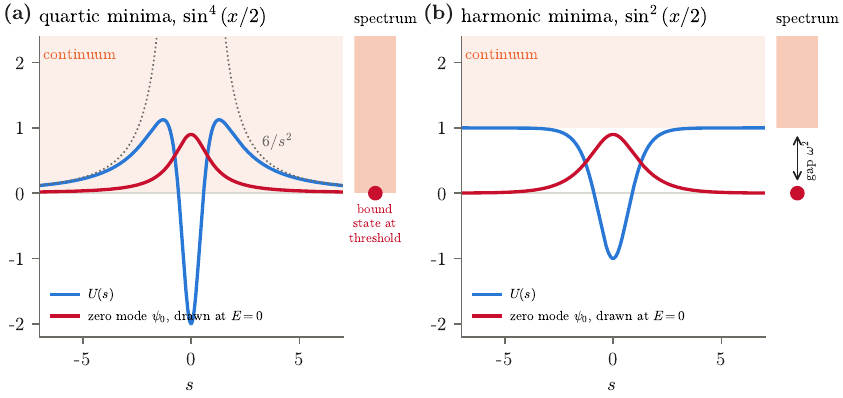}
\caption{The fluctuation operator and its spectrum. (a) Quartic 
minima: $U(s)=(6s^2-2)/(1+s^2)^2$ with tail $6/s^2$, and the 
zero mode $\psi_0\propto1/(1+s^2)$, drawn at $E=0$. The continuum
(shaded) starts at $0$, so the zero mode is a normalizable bound 
state at threshold. (b) Harmonic minima (sine-Gordon kink):
$U=1-2\sech^2s$, zero mode $\sech s$, and a gap up to $\omega^2=1$. 
The strips on the right show the spectra.}
\label{fig:fluct}
\end{figure}

Expanding $x=\xcl+\eta$ gives $S_E=S_0+\tfrac12\int\dd\tau\,
\eta\,M\,\eta+O(\eta^3)$ with $M=-\partial_\tau^2+V''(\xcl(\tau))$.
Using $V''=4G\big(12\sin^2\tfrac x2\cos^2\tfrac x2-4\sin^4
\tfrac x2\big)$, $\sin^2(\xcl/2)=1/(1+s^2)$ and $\cos^2(\xcl/2)
=s^2/(1+s^2)$,
\begin{equation}
M=\kappa^2\Big[-\partial_s^2+U(s)\Big],\qquad 
U(s)=\frac{6s^2-2}{(1+s^2)^2}.
\label{eq:U}
\end{equation}
This potential has the shape of a volcano (\cref{fig:fluct}a): 
a well $U(0)=-2$ near the core, zeros at $s=\pm1/\sqrt3$, a 
rim $U_{\max}=9/8$ at $s=\pm\sqrt{5/3}$, and a tail
\begin{equation}
U(s) = \frac{6}{s^2}-\frac{14}{s^4}+\dots
= \frac{l(l+1)}{s^2}+\dots,\qquad l=2 .
\end{equation}
Since $U\to0$, the continuum starts at zero: \emph{there is no gap}. 
The zero-energy solutions are nevertheless elementary. Translational
invariance gives the zero mode $y_1\propto\dd\xcl/\dd s$, and 
reduction of order gives the second solution,
\begin{equation}
y_1(s)=\frac{1}{1+s^2},\qquad 
\chi(s)=y_1(s)\int_0^s\frac{\dd s'}{y_1(s')^2}
  =\frac{s+\frac23s^3+\frac15s^5}{1+s^2},\qquad 
W[y_1,\chi]=1,
\label{eq:y1chi}
\end{equation}
with $y_1\simeq s^{-2}$ and $\chi\simeq s^3/5$ at large $|s|$ 
(the two solutions of the $l=2$ tail, $s^{-l}$ and $s^{l+1}$).
Everything below follows from these two functions.

The zero mode has no nodes, so by the Sturm oscillation theorem 
it is the ground state of $-\partial_s^2+U$: $M\ge0$, the instanton 
has no negative mode, and $y_1$ is the \emph{only} normalizable 
state, since any other bound state would have to lie below zero.
This also follows from $M/\kappa^2=A^\dagger A$ with $A=\partial_s
-y_1'/y_1$. What is unusual is \emph{where} this bound state is
located: exactly at the threshold of the continuum (\cref{fig:fluct}a).
In one dimension a state at threshold is normally a ``half-bound''
state, which tends to a constant at large $|s|$ and is not 
normalizable. Here the $l=2$ centrifugal tail makes the zero mode 
decay as $1/s^2$, and it is normalizable. For harmonic minima the 
zero mode is an isolated state, separated by a gap $\omega^2$
from the continuum (\cref{fig:fluct}b). Normalizable zero modes 
at the bottom of a massless continuum are also what one meets 
for the Yang--Mills instanton (\cref{sec:YM}).

\subsection{Gelfand--Yaglom determinants}
Having established the fluctuation operator and its zero mode, we now 
turn to the fluctuation determinant. Since the spectrum is gapless, with 
the zero mode lying at the threshold of the continuum, we work on a 
finite Euclidean-time interval and use the Gelfand--Yaglom method 
\cite{GY} to evaluate the determinant.

We take the strip $s\in[-L,L]$ with Dirichlet boundary conditions ($L$ 
in $s$ units) and start with normalizing the determinant by the determinant 
of the free operator. The Gelfand--Yaglom theorem \cite{GY} gives the
determinant of $M$ from the solution of $-y''+Uy=0$ with $y(-L)=0$, 
$y'(-L)=1$, as $\det M=y(L)$. The solution $y(s)$ has the following 
closed form: $y(s)=y_1(-L)\chi(s)-\chi(-L)y_1(s)$. The same theorem 
implies that
$\det(-\partial_s^2)_L=2L$,
\begin{equation}
\frac{\det M_L}{\det(-\partial_s^2)_L}
 = \frac{y_1(L)\chi(L)}{L}
 = \frac{1+\frac23L^2+\frac15L^4}{(1+L^2)^2}
 \;\xrightarrow{L\to\infty}\;\frac15 .
\label{eq:GY}
\end{equation}
At finite $L$ the zero mode becomes the lowest Dirichlet eigenvalue 
$\lambda_0(L)$. Writing $y_E$ for the solution of $-y_{E}''+Uy_{E}=Ey_{E}$
with $y_{E}(-L)=0$ and $y_{E}'(-L)=1$, we have $\det'M_L=\det M_L/\lambda_0
=-\partial_Ey_E(L)|_{E=0}$ up to relative corrections of order $\lambda_0/
\lambda_1$. Variation of parameters gives $\partial_Ey_E$ in closed form:
\begin{equation}
\det{}'M_L=\int_{-L}^{L}\dd s\,
   \Big[\chi(L)^2\,y_1(s)^2-y_1(L)^2\,\chi(s)^2\Big].
\label{eq:detprime}
\end{equation}
The integrals are elementary, and at large $L$
\begin{equation}
\frac{\det M_L}{\det(-\partial_s^2)_L}
     \;\simeq \;\frac15,\qquad
\frac{\det'M_L}{\det(-\partial_s^2)_L}
     \simeq\frac{\pi L^5}{100},\qquad 
\lambda_0(L)\simeq\frac{20}{\pi L^5},
\label{eq:detprime_asym}
\end{equation}
where the prime indicates that the lowest Dirichlet eigenvalue $\lambda_0$ 
is removed. The zero-mode-removed determinant \emph{diverges} as $L^5$. 
On the window the spectrum is discrete: the zero mode is lifted to 
$\lambda_0\simeq20/(\pi L^5)$, and the next levels, which become the
continuum as $L\to\infty$, are spaced as $L^{-2}$. The lifted zero mode 
is therefore separated from them only by a power of $L$, whereas for 
harmonic minima $\lambda_0\propto\e^{-2\omega L}$ and the non-zero
levels start above the gap $\omega^2$.
 
A related object with no zero mode at all is the Dirichlet 
determinant of \emph{half} the window, from the matching point 
to the midpoint. Place the midpoint (the barrier top $x_m=\pi$) 
at $\tau=0$; the instanton passes a point $x_a$ at $\tau=-\tau_a$, 
so $\tau_a>0$ is the time it needs to go from $x_a$ to the top
of the barrier. On $[-\tau_a,0]$ with Dirichlet conditions, 
$\Dh$ is the solution of the Jacobi equation $-\ddot y+V''(
x_{\rm cl})\,y=0$ with $y(-\tau_a)=0$, $\dot y(-\tau_a)=1$, 
evaluated at $\tau=0$. Reduction of order from the zero mode 
$\dot x$ gives this solution in closed form, $y(\tau)=
\dot x(\tau)\,\dot x_a\int_{-\tau_a}^{\tau}\dd\tau'/\dot x^2$, 
so
\begin{equation}
\Dh(L)=\dot x_a\,\dot x_m \int_{-\tau_a}^{0}
    \frac{\dd\tau}{\dot x^2} \qquad 
  \Big(=\chi(L)/\kappa\ \text{with $\kappa\tau_a=L$}\Big),
\label{eq:Dhalf}
\end{equation}
where $\dot x_m=2\kappa$ is the velocity at the top of the barrier.
Equivalently, $\Dh=\partial x_m/\partial p_a$ at fixed $\tau_a$, 
which is the classical Van Vleck form.\footnote{This relation can be 
seen as follows. Consider the classical solution $x_{cl}(\tau;p_{a})$ 
which obeys $\ddot{x}_{cl}=V'(x_{cl})$ with $x_{cl}(-\tau_{a};p_{a})
=x_{a}$ and $\dot{x}(-\tau_{a};p_{a})=p_{a}$. Define $J(\tau)\equiv
\frac{\partial x_{cl}(\tau;p_{a})}{\partial p_{a}}$. The function 
$J(\tau)$ obeys the equation $-\ddot{J}(\tau)+V''(x_{cl})J(\tau)=0$ 
with $J(-\tau_{a})=0$ and $\dot{J}(-\tau_{a})=1$. Therefore, the 
Gelfand-Yaglom theorem implies that $D_{\text{half}}=J(0)$.} 
It is finite and, as we will see, it is the object that appears 
naturally in the correct calculation of the tunneling rate. It 
is related to the full-window determinant by an identity that 
follows from \eqref{eq:detprime}. At large $L$ the first term 
in this expression dominates, and equals $\frac\pi2\chi(L)^2$. With 
$\Dh=\chi(L)/\kappa$, $S_0=2\pi\kappa$ and $\dot x_m=2\kappa$, 
converting to units of $\tau$,
\begin{equation}
\det{}'M_L=\frac{\pi\chi(L)^2}{2\kappa^3}(1+O(L^{-3}))
  =S_0\,\frac{\Dh(L)^2}{\dot x_m^2}\,
    \big[ 1+O(L^{-3}) \big].
\label{eq:identity}
\end{equation}

\subsection{The zero mode as a flux through the top of the  barrier}

 Time translation $\tau_0\to\tau_0+\epsilon$ changes the instanton 
by $\epsilon\,\dot x_{\rm cl}$, which is annihilated by the 
operator $M$. With $\int\dot x_{\rm cl}^2\dd\tau=S_0$ the 
normalized zero mode is
\begin{equation}
\psi_0(\tau) = \frac{\dot x_{\rm cl}(\tau)}{\sqrt{S_0}}
 \quad \Rightarrow \quad \psi_0(s)
=\sqrt{\frac{2}{\pi}}\,\frac{1}{1+s^2},
\end{equation}
where $\psi_0(s)$ is normalized as a function of $s$.
The Gaussian integral along this direction is replaced by an 
integral over the collective coordinate $\tau_0$ with Jacobian 
$\sqrt{S_0/2\pi\hbar}$. The $\tau_0$ integral produces the factor 
$T$ (total Euclidean time) that turns the tunneling rate $t$ 
into an energy shift. The zero mode is normalizable, but it is 
a bound state at threshold. On a strip it is lifted only by a 
power, $\lambda_0\propto L^{-5}$.

Combining the Jacobian with the lifted eigenvalue and using 
\eqref{eq:identity} gives
\begin{equation}
\sqrt{\frac{S_0}{2\pi}}\,\big[\det{}'M_L\big]^{-1/2}
  = \frac{\dot x_m}{\sqrt{2\pi}\,\Dh(L)}.
\label{eq:flux}
\end{equation}
The factor $\sqrt{S_0}$ from the zero-mode Jacobian cancels 
against the norm of the lifted zero mode. What is left is the
\emph{velocity at the top of the barrier} divided by a Dirichlet
determinant with no zero mode. In other words, integrating over 
$\tau_0$ is the same as integrating over the time at which the 
path crosses the top of the barrier, weighted by the flux $\dot x_m$. 
This is the form in which the collective coordinate will be used 
in \cref{sec:matching}: the zero mode is traded for a probability
current through the barrier top, as in Bardeen's transfer-Hamiltonian
picture of tunneling \cite{Bardeen}, and as in the exact amplitude 
formula of \cref{sec:herring}.

\subsection{Failure of the textbook formula}

 The standard procedure is to insert \eqref{eq:detprime_asym} 
into \eqref{eq:coleman} with the free operator $-\partial_\tau^2$ 
as reference (this is the Gaussian approximation for a potential 
that satisfies $V''(0)=0$). With $L_\tau=L/\kappa$ the window 
half-width in $\tau$ units,
\begin{equation}
K(L) = \sqrt{\frac{S_0}{2\pi}}
   \left[\frac{\det'M_L}{\det(-\partial_\tau^2)_L}\right]^{-1/2}
     = \sqrt{\frac{L_\tau}{\pi}}\,\frac{\dot x_m}{\Dh(L)}
     = \frac{10\,\kappa^{3/2}}{\sqrt\pi\,L^{5/2}}\;
     \xrightarrow{L\to\infty}\;0.
\label{eq:Knaive}
\end{equation}
Thus, with quartic minima, \eqref{eq:coleman} with a free reference
operator depends on the arbitrary window size and vanishes as 
$L^{-5/2}$. The culprit is the reference: $-\partial_\tau^2$ 
describes a free particle, which is the Gaussian approximation 
to a vacuum that is not Gaussian at all. For harmonic minima, 
$\det(-\partial^2+\omega^2)_L\propto\e^{2\omega L}$ grows exactly 
like $\det'M_L$, and the ratio gives the finite, standard answer 
$K=A\sqrt{\omega/\pi}$ with $\dot x\simeq A\,\e^{-\omega|\tau|}$.

It is worth emphasizing that the denominator of \eqref{eq:coleman} 
is not a convention. It is the vacuum sector of the same path 
integral, evaluated by the same Gaussian method, and dividing by 
it is what turns the instanton contribution into an energy shift. 
For a harmonic minimum the Gaussian vacuum amplitude is the correct
vacuum amplitude. For $V''(0)=0$ the Gaussian approximation to the
vacuum sector is a free particle, which has no normalizable ground 
state and is not an approximation to the true vacuum at all. No 
Gaussian reference can repair this: $\det'M_L$ grows as $L^6$ in 
$\tau$ units, the free determinant supplies one power of $L$, and a 
massive one grows exponentially. The reference that does the job 
is the exact vacuum amplitude, which is what \cref{sec:matching} 
uses.

\section{Instanton interactions}
\label{sec:interactions}

The interaction of two tunneling events separated by a Euclidean 
time $T$ is the action of the two-event configuration minus $2S_0$. 
Two channels matter: an instanton followed by an anti-instanton
($I\Ib$), which returns to the starting vacuum, and two instantons
($II$), which wind twice. In this section we explain why linearly 
superposing two instantons fails for quartic minima, and derive the 
correct classical formula $S_{\rm int}\propto T^{-3}$ from the 
separatrix of the Euclidean motion. Finally, we explain what replaces 
the classical result beyond the quantum time $\tau_q$.

\subsection{Failure of superposition for quartic minima}

For harmonic minima the equation of motion is nonlinear only 
inside the instanton cores. Outside them the deviation from the 
vacuum obeys the \emph{linear} equation $\ddot\eta=\omega^2\eta$, so 
two well-separated instantons can be linearly superposed. The sum 
solves the equations everywhere except in the cores, and the 
interaction is the first-order (boundary) term, $S_{\rm int}\propto
\mp \e^{-\omega T}$ \cite{Coleman,ZJ}. For quartic minima, there is 
no linear region at any distance: the tails obey $\ddot\eta=
4G\eta^3$ all the way to the vacuum. It is instructive to see 
what the sum gives anyway. Take
\begin{equation}
x_{II}  = 2\pi+2\arctan s+2\arctan(s-T),\qquad 
x_{I\Ib}= 2\arctan s-2\arctan(s-T)\, ,
\end{equation}
where $T$ is given in the same units as $s$. The integrand of the 
action is a rational function of $s$, and the integrals are 
elementary:
\begin{equation}
\frac{S[x_{II}]-2S_0}{S_0}=\frac{4}{T^2+4},\qquad 
\frac{S[x_{I\Ib}]-2S_0}{S_0}=\frac{4\,(T^4-20T^2-32)}{(T^2+4)^3}.
\label{eq:sumansatz}
\end{equation}
Both tend to $+4/T^2$. The sign is the same (repulsive) in both
channels, and the power is wrong: the exact results below are 
$+20.1/T^3$ and $-5.0/T^3$ in the same units. Both expressions 
depend only on $T^2$, so no $T^{-3}$ term can appear at any 
order.

Why does the sum fail so badly? Near the core of the first instanton
the tail of the second instanton is an almost constant shift, 
$\eta\simeq\pm2/T$. A constant is not a zero mode of $M$, because
$U(s)\ne0$ in the core. The sum therefore pays a second-order cost
\begin{equation}
\frac{\kappa}{2}\int\dd s\,\big[\dot{\eta}^2+U\eta^2\big]
 \simeq \frac{\kappa}{2}\,\frac{4}{T^2}\int U\,\dd s
  = \frac{2S_0}{T^2}\qquad\text{per core},
\end{equation}
which is exactly the $4S_0/T^2$ of \eqref{eq:sumansatz}. It is 
positive because $M\ge0$, which is why both channels look
repulsive.

 The problem with this calculation becomes apparent when 
we compare the error of the ansatz with the size of the 
effect we are trying to compute,
\begin{equation}
\text{quartic:} \quad 
    \underbrace{\frac{S_0}{T^2}}_{\rm error}
 \;\gg\;
    \underbrace{\frac{S_0}{T^3}}_{\rm signal},\qquad\qquad
\text{harmonic:}\quad 
    \underbrace{S_0\,\e^{-2\omega T}}_{\rm error}
\;\ll\;
    \underbrace{S_0\,\e^{-\omega T}}_{\rm signal}.
\end{equation}
For exponential tails the error is second order in the 
overlap and harmless, but for power-law tails it is larger than 
the signal by a factor $T$. The sum is still a legitimate 
trial configuration, so it bounds the constrained action 
from above, but the bound is not tight.

\subsection{The exact interaction from the separatrix}
\label{sec:separatrix}

The correct interaction can be obtained from the exact classical
solution, and the simplest route to it is the one-dimensional 
version of Manton's argument for the forces between kinks
\cite{Manton,Christov}: the force between two kinks can be 
computed using an exact static solution between the two objects, 
and then determining the force from the stress at the location
of the kinks. This calculation can be organized using the 
separatrix of the phase plane. Here we explain how to apply
this idea to multi-instanton events in the Euclidean phase
plane shown in \cref{fig:separatrix}. The procedure is as
follows:

\begin{figure}[!t]
\centering
\includegraphics[width=0.92\textwidth]{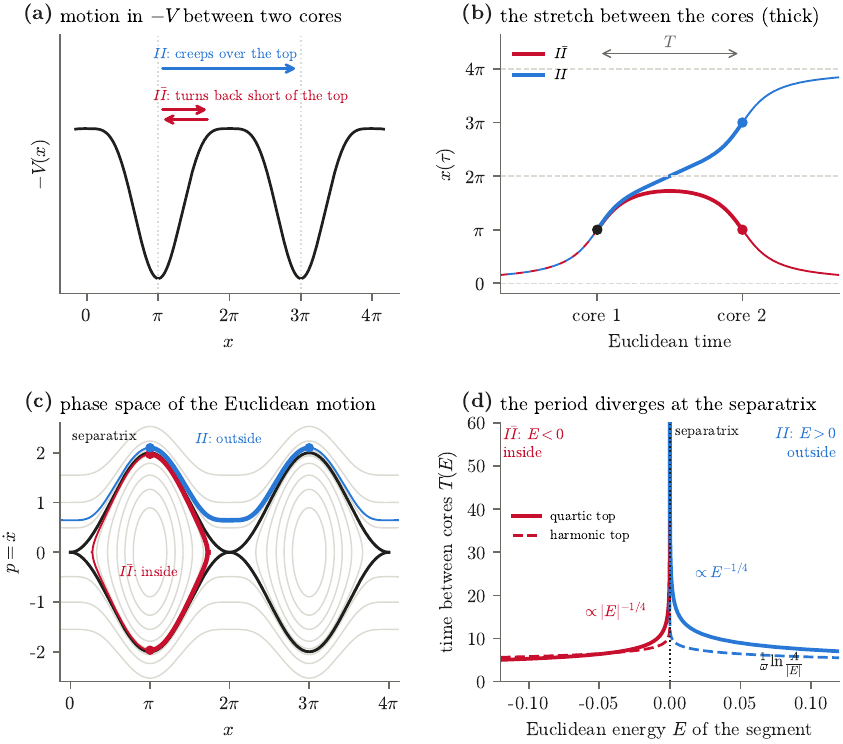}
\caption{The separatrix derivation ($\kappa=1$). (a) In the 
inverted potential $-V$, the stretch between two cores hovers 
near a hilltop. For $I\Ib$ it turns back short of the top; for 
$II$ it creeps over it. (b) The same stretches in Euclidean time, 
for $T=6$. Coloured: the stretch between the cores, which carries 
the interaction. Black and dashed: the outer halves of the two
events. (c) Phase plane of the Euclidean motion, with the 
separatrix $p=\pm\sqrt{2V}$ (black). The $I\Ib$ stretch (red) 
lies inside it and returns to its starting core, and the $II$ 
stretch (blue) lies outside and passes to the next core. (d) The 
time between cores as a function of the Euclidean energy $E$ of the
stretch. For a quartic top ($\sin^4(x/2)$, solid) it diverges as 
$|E|^{-1/4}$ on both sides, and the $II$ side lies higher by 
$\sqrt2$. For a harmonic top ($\sin^2(x/2)$, dashed) the 
divergence is logarithmic and the two sides coincide.}
\label{fig:separatrix}
\end{figure}

\paragraph{Step 1: cut at the cores.}
Cut the two-event history at the two cores, the points $x=\pi$ 
(mod $2\pi$) where each event is half-way through
(\cref{fig:separatrix}a,b). The colored interpolation (``stretch'')
between the cores is the only part of the history that knows 
about $T$. Consider an exact classical trajectory describing a repeating
sequence of instanton-like events. Each segment between neighboring cores 
has the same duration $T$ and action $S_{\rm seg}(T)$. For $T\to\infty$ 
the stretch becomes the second half of one instanton followed by the 
first half of the next event. Each half costs $S_0/2$, so
\begin{equation}
S_{\rm int}(T)=S_{\rm seg}(T)-S_0 .
\end{equation}

\paragraph{Step 2: the separatrix.}
The Euclidean motion conserves $E=\tfrac12\dot x^2-V$, so it is 
a motion in the phase plane (\cref{fig:separatrix}c). An isolated
instanton has $E=0$. It runs along the \emph{separatrix}
$p = \pm\sqrt{2V}$, which connects the hilltops of $-V$ and needs
infinite time to reach them. Orbits with $E<0$ lie inside the
separatrix: they turn back before reaching the hilltop and return 
to the core they came from. That is the $I\Ib$ channel. Orbits with
$E>0$ lie outside: they creep over the hilltop and go on to the 
next core. That is the $II$ channel. So the sign of $E$ labels 
the channel, and the separatrix is the dividing line between 
them\footnote{The instanton equation has a first-order form, 
$\dot x=W'(x)$ with $W'=\sqrt{2V}=2\kappa\sin^2(x/2)$ and
$W(x)=\kappa(x-\sin x)$, the same function that appears in
\cref{sec:matching}. Because $W'\ge0$, every right-moving hop, 
in every cell, solves the same first-order equation, and the
Bogomolny bound $S\ge|\Delta W|=wS_0$ holds for winding $w$. The 
$II$ configuration is therefore asymptotically BPS: its action
approaches $2S_0$ from above as $T\to\infty$, and the repulsion
\eqref{eq:interactions} measures the departure from the bound. 
For harmonic minima, $W'\propto\sin(x/2)$ changes sign from cell 
to cell, the second right-moving hop solves the first-order equation
with the opposite sign. Therefore, the net change in $W$ then vanishes, 
and the Bogomolny bound reduces to the trivial inequality $S\ge0$. What 
a quartic minimum does not allow is exact gluing: $W'$ vanishes
quadratically at the vacuum, so the flow reaches it only as 
$\tau\to\infty$ and two hops at finite separation never solve 
the first-order equation. The exact $II$ solution is the rotating
orbit with $E>0$.}.

\paragraph{Step 3: the force is the energy.}
The stretch is a classical path from one core to the next in 
the time $T$. By the Hamilton--Jacobi relation, the derivative of 
its Euclidean action with respect to the elapsed time, at fixed 
end points, is minus its energy $E=\tfrac12\dot x^2-V$. Hence
\begin{equation}
\frac{\dd S_{\rm int}}{\dd T}
  = -E,\qquad S_{\rm int}(T)
  = \int_T^\infty E(T')\,\dd T' ,
\label{eq:forceE}
\end{equation}
where we used $S_{\rm int}\to0$ as $T\to\infty$. The conserved 
Euclidean energy is the force conjugate to the separation, $E=-
dS_{\rm int}/dT$: negative $E$ corresponds to attraction, while 
positive $E$ corresponds to repulsion. This is the one-dimensional 
analogue of Manton's stress-tensor argument: in $1+1$ dimensions 
the force on a kink is the flux of the stress tensor through a 
point between the kinks. In Euclidean quantum mechanics the 
stress tensor is the single number $\tfrac12 \dot x^2-V$, 
conserved along the stretch. Everything reduces to the function 
$E(T)$, i.e.\ to the time $T(E)$ the orbit needs to go from core 
to core.

\paragraph{Step 4: the period near the separatrix.}
The time spent hovering diverges as $E\to0$ (\cref{fig:separatrix}d),
and how it diverges depends on the shape of the hilltop:

\begin{itemize}

\item \emph{Quartic hilltop}, $V\simeq G\eta^4$ with $\eta$ the distance 
from the top and $\kq=\sqrt{2G}$. For $E<0$, the orbit turns at $G\eta_t^4
=|E|$. For $E>0$, it crosses the hilltop with nonzero velocity; we define 
the characteristic crossing scale $\epsilon=(E/G)^{1/4}$. Then
\begin{align}
T_{I\Ib} &= T_0 + \frac{2}{\kq\eta_t}\int_1^\infty
  \frac{\dd u}{\sqrt{u^4-1}}
          = T_0 + \frac{c}{\kq\,\eta_t},\nonumber\\
T_{II}   &= T_0 + \frac{1}{\kq\epsilon}\int_{-\infty}^{\infty}
  \frac{\dd v}{\sqrt{1+v^4}}
          = T_0 + \frac{\sqrt2\,c}{\kq\,\epsilon},
\label{eq:Tsep4}
\end{align}
with
\begin{equation}
c=\frac{\Gamma(\frac14)^2}{2\sqrt{2\pi}}=2.6221 .
\end{equation}
Both diverge as $|E|^{-1/4}$, but the crossing takes longer 
than the reflection by exactly $\sqrt2$. For $\sin^4(x/2)$ 
the regular offset vanishes, $T_0=0$, because the separatrix 
time from the core to a distance $\eta$ from the top, 
$\cot(\eta/2)/\kappa=1/(\kq\eta)-\eta/(6\kappa)+\dots$, has 
no constant term.

\item \emph{Harmonic hilltop}, $V\simeq\frac12\omega^2\eta^2$.
Reflection gives an $\operatorname{arccosh}$ and crossing an 
$\operatorname{arcsinh}$. For large arguments both behave as 
$\ln(2y)$, with the same constant, so the two sides of the 
separatrix give the same time:
\begin{equation}
T_{I\Ib}=T_{II}=\frac1\omega\ln\frac{A}{|E|}+o(1),\qquad
 A=2\omega^2A_t^2,
\label{eq:Tsep2}
\end{equation}
where $A_t$ is the tail amplitude, $\eta\simeq A_t\,
\e^{-\omega|\tau|}$. The coefficient of the logarithm is 
$1/\omega$ on both sides.
\end{itemize}

\paragraph{Step 5: integrate the force.}
\begin{itemize}
\item \emph{Quartic.} Inverting \eqref{eq:Tsep4} and using 
$\kq^4=4G^2$, $|E|=c^4/[4G(T-T_0)^4]$ for $I\Ib$ and 
$E=c^4/[G(T-T_0)^4]$ for $II$. Then \eqref{eq:forceE} 
gives
\begin{equation}
S_{I\Ib}= - \frac{c^4}{12\,G\,(T-T_0)^3},\qquad 
S_{II}  = + \frac{(\sqrt2)^4c^4}{12\,G\,(T-T_0)^3}
        = + \frac{c^4}{3\,G\,(T-T_0)^3}.
\label{eq:interactions0}
\end{equation}
The factor $4$ between the channels is $(\sqrt2)^4$: the ratio 
of crossing to reflection time, raised to the fourth power, 
because $E\propto(T-T_0)^{-4}$.

\item \emph{Harmonic.} Inverting \eqref{eq:Tsep2}, $E=\mp A\,
\e^{-\omega T}$, and
\begin{equation}
S_{\rm int} = \mp\frac{A}{\omega}\,\e^{-\omega T}
            = \mp\,2\omega A_t^2\,\e^{-\omega T}.
\end{equation}
This is the standard result, and the two channels have equal 
strength because the logarithm has the same coefficient on both 
sides.
\end{itemize}
Since $G=\tau_q^{-3}$, the quartic result is universal in 
units of the vacuum time:
\begin{equation}
\frac{S_{I\Ib}}{\hbar} \simeq
  -3.94\left(\frac{\tau_q}{T}\right)^3,\qquad 
\frac{S_{II}}{\hbar} \simeq
  +15.76\left(\frac{\tau_q}{T}\right)^3.
\label{eq:interactions}
\end{equation}
The classical interaction becomes $O(\hbar)$ precisely at
$T\sim\tau_q$, where the classical tails dissolve into 
the zero-point motion. Computing $E(T)$ for the exact periodic 
orbits by quadrature with $T(E)=\int_{\text{seg}} \frac{dx}{\sqrt{2V(x)+E}}$ reproduces both quartic results to $0.2\%$ 
accuracy at $|E|=10^{-5}V_{\max}$. The orbits reach the law 
\eqref{eq:interactions} once $T\gtrsim10\,\tau_I$. Integrating
once more, \eqref{eq:forceE} can be written through the abbreviated
action of the stretch, $W(E)=\int_{\rm seg}p\,\dd x$ with
$p=\sqrt{2(V+E)}$, as the Legendre transform
\begin{equation}
S_{\rm int} = W(E)-W(0)-E\,T(E),\qquad 
 T(E)=W'(E).
\label{eq:legendre}
\end{equation}
Any part of $W(E)-W(0)$ that is analytic at the separatrix drops 
out at leading order, and only the non-analytic part, which comes 
from the solution hovering near the hilltop, survives. This form 
reappears in Appendix~\ref{app:spectral}.

The interaction is therefore fixed by the singularity of the 
period at the separatrix. A logarithmic divergence of the period,
$T\simeq\omega^{-1}\ln(1/|E|)$, gives an exponential interaction 
with rate $\omega$. The coefficient of the logarithm is the 
Lyapunov exponent of the hilltop of $-V$, which equals the gap of 
the harmonic well. It is the same on both sides of the separatrix, 
so $|S_{II}|=|S_{I\Ib}|$. A power-law divergence, $T\propto
|E|^{-1/4}$, gives a power-law interaction. Its coefficients on 
the two sides differ by the ratio of crossing to reflection time, 
$\sqrt2$, so $|S_{II}|=4|S_{I\Ib}|$. The sign is the side of the
separatrix: inside ($I\Ib$, attraction) or outside ($II$, repulsion).

The power three has a simple origin. The tail equation $\ddot\eta
= 4G\eta^3$ is invariant under $\eta\to\lambda\eta$, $\tau\to
\tau/\lambda$. The tail action $\int\dd\tau\,(\frac12\dot\eta^2 
+ G\eta^4)$ scales as $\lambda^3$ and the separation as $1/\lambda$, 
so $S_{\rm int}\propto T^{-3}$; only the constant needs the nonlinear
solution. The linear tail equation of harmonic minima has a built-in
time $1/\omega$ and gives an exponential; the quartic one is scale
free and gives a power law. Since $|E|=V(\eta_t)$, the $I\Ib$ result
can also be written as $S_{I\Ib}=-\tfrac13\,T\,V(\eta_{\rm mid})$: 
the interaction is one third of the potential at the midpoint 
times the time spent there.

\subsection{Beyond the vacuum time: exchange of a gapped excitation}
\label{sec:beyondtauq}

The classical tails make sense only when the motion is outside 
the quantum width, $x\gg\ell$, i.e.\ for $T\ll\tau_q$. The 
classical interaction therefore requires
\begin{equation}
\tau_I\ll T\ll\tau_q,\qquad\frac{\tau_q}{\tau_I}=2\sqrt2\,G^{1/6},
\end{equation}
a window of about $3$ at $S_0=22$ and $14$ at $S_0=2200$.

What happens for $T\gg\tau_q$ is clearest in the Hamiltonian 
picture. Between the two events the path sits in the well of the non-inverted potential for a time 
$T$, and the path integral over that stretch is the propagator of 
the well. Expanding it in the eigenstates of the well, with the
stretch cut at the two cores $\xi_{1,2}$ (Appendix~\ref{app:spectral}),
\begin{equation}
\e^{-S_{\rm int}(T)/\hbar} 
   =  \sum_{n\ge0}R_n\,\e^{-(E_n-E_0)T/\hbar},\qquad 
R_n=\frac{\psi_n(\xi_2)\,\psi_n(\xi_1)}
         {\psi_0(\xi_2)\,\psi_0(\xi_1)} .
\label{eq:spectral}
\end{equation}
The dilute gas keeps only $n=0$, and the interaction is the rest. 
For $T\gg\tau_q$ the first excited state dominates:
\begin{equation}
S_{\rm int}(T) 
  \simeq  -\,r_1\,\e^{-(e_1-e_0)T/\tau_q}
  =  -\,r_1\,\e^{-1.726\,T/\tau_q}\qquad(T\gg\tau_q).
\label{eq:exchange}
\end{equation}
The first excited state is odd about the center of the well. 
For $I\Ib$ the path enters and leaves through the same barrier, 
so $r_1>0$ (attraction). For $II$ it leaves through the opposite
side, so $r_1<0$ (repulsion). These are the same signs as in the
classical regime. Since $r_1=O(1)$, this quantum interaction carries
no factor $1/\hbar$. Appendix~\ref{app:spectral} shows that the same sum 
also contains the classical regime: for $T\ll\tau_q$ it is dominated
by a band of highly excited states, and its saddle point is exactly
the Legendre transform \eqref{eq:legendre}.

Instantons therefore interact by exchanging vacuum excitations. 
Classically the vacuum is gapless, the exchange is ``massless'', 
and the interaction is a power law, \eqref{eq:interactions}. 
Quantum mechanically the vacuum has a dynamically generated gap
$1.726\,E_q$. For $T\gg\tau_q$ the exchange is massive and the
interaction is exponential, \eqref{eq:exchange}. For harmonic 
minima the gap is $\omega$, the same rate as the classical tails, 
and classical and quantum formulas agree (Appendix~\ref{app:spectral}). This is 
why the textbook case never shows the distinction.

For the dilute gas, typical separations are $\bar T\sim1/t
\simeq\tau_q\,\e^{S_0}/1.5\gg\tau_q$, so the non-interacting gas picture is safe at 
leading order. Close pairs, which matter at order $\e^{-2S_0}$
\cite{ZJJ2004,DunneUnsal}, are dominated by separations 
$T\sim\tau_q$, exactly the crossover, and carry no large 
logarithm $\ln(S_0/\hbar)$ as in the harmonic case.

\section{Quantum fluctuations: the bottom of the well is quartic}
\label{sec:vacuum}

Inside the well, $V=16G\sin^4\frac x2=G\big(x^4-\tfrac16x^6+
\tfrac1{80}x^8-\dots\big)$. With $x=\ell y$,
\begin{equation}
H = E_q\Big[-\tfrac12\partial_y^2
  +  y^4-\tfrac16\,\ell^2y^6+\tfrac1{80}\,\ell^4y^8-\dots\Big],\qquad
   \ell^2=G^{-1/3}.
\label{eq:Hwell}
\end{equation}
The leading operator $-\tfrac12\partial_y^2+y^4$ contains no 
parameter at all: the coupling has been scaled out. Low lying 
dimensionless energies $e_i$ and ground state expectation values 
$\langle y^{2n} \rangle$ for the leading operator can be computed numerically 
(Appendix~\ref{app:numerics}, see also \cite{TurbinerDelValle,SSV}). 
We obtain
\begin{align}
\label{eq:quarticconstants}
e_0&=0.667986,\quad 
e_1=2.393644,\quad 
e_2=4.696795,\qquad  \\
\langle y^2\rangle &=0.287338,\quad 
\langle y^6\rangle=0.265163 .  \nonumber
\end{align}
The asymptotic normalization $B$ of the ground state 
(for $y>0$) is 
\begin{equation}
\varphi_0(y)\simeq
  \frac{B}{y}\,\exp\!
    \Big(-\frac{\sqrt2}{3}y^3-\frac{e_0}{\sqrt2\,y}\Big),\qquad
    B=1.028286 .
\label{eq:B}
\end{equation}
We note that the value of $\langle y^6\rangle$ can be checked 
using the hypervirial identity $10\langle y^6\rangle=6e_0
\langle y^2\rangle+\tfrac32$. Perturbation theory in the well
therefore involves powers of $\ell^2\propto\hbar^{2/3}$, not 
$\hbar$. First-order perturbation theory in the $x^6$ term 
gives
\begin{equation}
E_0 = e_0\,G^{1/3} - \tfrac16\langle y^6\rangle + O(G^{-1/3})
  =0.667986\,G^{1/3}-0.04419-0.0020\,G^{-1/3}+\dots,
\label{eq:E0series}
\end{equation}
where the last coefficient follows from second-order perturbation 
theory together with the $\tfrac1{80}\ell^4y^8$ term. The classical
vacuum is gapless, but the quantum vacuum has a gap $(e_1-
e_0)E_q=1.726\,(\hbar^4\lambda/I^2)^{1/3}\propto\hbar^{4/3}$, 
which is an intrinsically quantum scale. The vacuum is ``quiet'',
$\ell\to0$ as $G\to\infty$, but not Gaussian at any coupling: 
its levels are spaced as $n^{4/3}$ and its tail is 
$\e^{-\sqrt2y^3/3}$. This is what the free reference operator 
in \eqref{eq:Knaive} gets wrong.

The classical tail $x\simeq1/(\kq|\tau|)$ enters the quantum 
region $x\lesssim\ell$ at
\begin{equation}
|\tau| \simeq \frac{1}{\kq\ell} = \frac{\tau_q}{\sqrt2}.
\end{equation}
Beyond that time the path is zero-point motion in the quartic 
well, and the classical tail no longer exists. The resulting 
hierarchy of time scales is
\begin{equation}
\tau_I\;\ll\;\tau_q\;\ll\;\bar T,\qquad 
\frac{\tau_q}{\tau_I}=2\sqrt2\,G^{1/6},\qquad 
\frac{\bar T}{\tau_q}=\frac{E_q}{t}\simeq\frac{\e^{S_0}}{1.5}.
\end{equation}
The instanton gas is always exponentially dilute, but a single 
instanton is never much shorter in duration than the vacuum 
time: the first ratio grows only as $G^{1/6}$ (it is $3$ at
$S_0=22$). This is the origin of the power-law corrections 
found below, which are of order $(\tau_I/\tau_q)^2=\ell^2/8$.

The pure quartic vacuum is accurate near the minimum and fails 
under the barrier, because $16\sin^4(x/2)\ne x^4$. WKB is accurate 
under the barrier and fails near the minimum. They overlap in the
region
\begin{equation}
\ell\ll x\ll\ell^{3/5}\qquad
  \Longleftrightarrow \qquad 
  1\ll y\ll G^{1/15}.
\label{eq:overlap}
\end{equation}
The lower limit is where WKB starts to work and the vacuum energy
becomes negligible compared to the potential, $x\gg\ell$. The upper 
limit is where the pure $x^4$ vacuum stops working: the $x^6$ term
adds to the WKB exponent
\begin{equation}
\int_0^x\sqrt{2V}=\kq\Big(\frac{x^3}{3}-\frac{x^5}{60}+\dots\Big),
\label{eq:Wexp}
\end{equation}
whose second term is small only for $\kq x^5\ll1$, i.e.\ 
$x\ll\ell^{3/5}$. The window opens very slowly, but the formal
matching organizes all corrections as a series in $\ell^2$.

 The instanton core and the vacuum require 
different treatments. Fluctuations around the core admit a semiclassical 
expansion controlled by $\hbar_{\rm eff}=\pi/S_0$. Near the minimum, 
however, the leading quantum problem is the parameter-free quartic 
oscillator, whose vacuum is non-Gaussian and has no small parameter for a 
loop expansion. We therefore evaluate the core semiclassically and the 
vacuum quantum mechanically, matching the two descriptions in the overlap 
region \eqref{eq:overlap}. The vacuum contribution remains part of the 
path integral; it is its Gaussian approximation that fails. In quantum
mechanics, this contribution can be expressed directly through the 
vacuum wavefunction, whose leading behavior is fixed by the universal 
constants $e_0$ and $B$. The next two sections implement this 
matching.

\section{Making the instanton calculation match the spectrum}
\label{sec:matching}

 In this section we describe the correct one-instanton calculation. 
We begin with the exact answer, which follows from the Schr\"odinger
equation. We then derive the same answer from the path integral, 
which is where the instanton, its zero mode and its half-window 
determinants appear, and where one sees exactly which part of the 
textbook calculation has to be replaced. We then compute the first
correction, and finally we compare with the numerically computed 
spectrum.

\subsection{The transition amplitude from the wavefunction}
\label{sec:herring}

Let $\psi_L$ be the state localized in the well at $x=0$, with 
energy $\epsilon$, and $\psi_R(x)=\psi_L(2\pi-x)$ the corresponding
state in the adjacent well. We consider the two combinations 
$\psi_\pm =(\psi_L\pm\psi_R)/\sqrt2$, with energies $E_\pm$.
We can subtract the Schr\"odinger equations for $\psi_L$ (valid 
for $x\le\pi$) and $\psi_\pm$, and integrate up to the top of
the barrier
\begin{equation}
(E_\pm-\epsilon)\int_{-\infty}^{\pi}\psi_L\psi_\pm\,\dd x
 = -\frac{\hbar^2}{2I}
   \Big[\psi_L\psi_\pm'-\psi_\pm\psi_L'\Big]_{x=\pi}.
\end{equation}
At the top of the barrier $\psi_+'=0$ and $\psi_-=0$, and 
the overlap on the left is $1/\sqrt2$ up to exponentially 
small terms. This gives $E_\pm=\epsilon\mp t$ with
\begin{equation}
t=\frac{\hbar^2}{I}\,\psi_L(\pi)\,\big|\psi_L'(\pi)\big|
\label{eq:herring}
\end{equation}
The tunneling amplitude is the probability current that the 
single-well state tunnels through the barrier\footnote{The 
formula goes back to Herring \cite{Herring}, in the context of
exchange splittings in molecules; a version appears as a problem 
in Landau and Lifshitz \cite{LL}, and \cite{Garg} reviews its 
use for the one-dimensional double well.}. 
For a double well it gives the splitting $2t$. On the circle 
each cell is bounded by two barrier tops, and the same argument applied 
to the Bloch states at $\theta=0$ and $\pi$ gives the lowest 
band
\begin{equation}
E(\theta)=E_0-2t\cos\theta+O(\e^{-2S_0}),
\label{eq:band}
\end{equation}
with the same $t$ and $E_0=\epsilon$. Two features of
\eqref{eq:herring} matter for us. It is finite and never refers 
to a Gaussian vacuum: it requires only the single-well state near
the top of the barrier. And it is \emph{exact} to all orders in 
the expansion around the instanton: if $\psi_L$ is the all-orders
perturbative state of the well (a Bender--Wu expansion around the
quartic oscillator, continued under the barrier by WKB), then
\eqref{eq:herring} returns the all-orders one-instanton series, 
and the only error is of relative order $\e^{-2S_0}$. This is why it can serve as a benchmark for 
the path-integral calculation below, and later as the simplest 
way to compute corrections.

At leading order, under the barrier, $\psi_L=C\,p^{-1/2}\e^{-W(x)}$ 
with $p=\sqrt{2V}$ and $W(x)=\int_0^x\sqrt{2V}=\kappa(x-\sin x)$. 
This WKB form must match the quartic vacuum \eqref{eq:B} in the 
overlap region \eqref{eq:overlap}, where both are valid, and 
comparing $x^{-1}\e^{-\kq x^3/3}$ with $p^{-1/2}\e^{-W}\simeq
(\kq x^2)^{-1/2}\e^{-\kq x^3/3}$ fixes $C^2=\sqrt2B^2G^{1/3}$. 
At the top of the barrier $\psi_L'=-p\,\psi_L$ and $W(\pi)=S_0/2$, 
so $t=p\,\psi_L^2=C^2\e^{-S_0}$:
\begin{equation}
t= \sqrt2\,B^2\,G^{1/3}\,\e^{-4\pi\sqrt{2G}}
 = 1.4953\,E_q\,\e^{-S_0},\qquad 
   \sqrt2B^2=1.4953,
\label{eq:tleading}
\end{equation}
or, in physical units, $t=1.4953\,(\hbar^4\lambda/I^2)^{1/3}
\exp\!\big(-4\pi\sqrt{2I\lambda}/\hbar\big)$. Compared with 
the harmonic minima there is no factor $\sqrt{S_0/\hbar}$, and 
the prefactor scales as $\hbar^{4/3}$. It depends only on the 
quartic coefficient at the minimum and on $S_0$: the constants
governing the tail of the instanton have dropped out, because 
a quartic tail has no scale. The leading WKB current $p\,
\psi^2\e^{2W}$ is constant along the barrier, which is why 
it can be evaluated at the top of the barrier or anywhere 
in the overlap region.\footnote{In symmetric form the formula reads
$t=\frac{\hbar^2}{2I}\,|\psi_{L}\psi'_{R}-\psi_{R}\psi'_{L}|$,
with $\psi_{L}=C\,p^{-1/2}\e^{-W(x)}$ and $\psi_{R}=C\,
p^{-1/2}\e^{-(S_{0}-W(x))}$ under the barrier. The $p'$ terms cancel
in the Wronskian, which equals $2p\,\psi_{L}\psi_{R}=2C^2\e^{-S_0}$
at every point under the barrier, so $t=\hbar^2C^2\e^{-S_{0}}/I$.}

\subsection{The path integral in a gapped vacuum}
\label{sec:pathint}

The one-instanton calculation corresponds to a path integral
with one saddle point. The only thing wrong with the textbook
version is that it evaluates \emph{every} part of that path
integral semiclassically, including the long sections of the
path where the trajectory is close to a classical vacuum.
The cure is to cut the path where the semiclassical treatment
stops being valid, and to evaluate the two sections of the
paths using approaches that are reliable in each regime
(\cref{fig:picture}). Between the matching points
$x_a$ and $2\pi-x_a$, in the overlap region \eqref{eq:overlap},
the path is a half-instanton followed by another half-instanton,
with Gaussian fluctuations. The endpoints are fixed, so there
is no zero mode, and the determinants are the finite
half-window ones, $\Dh$ of \eqref{eq:Dhalf}. Outside the
matching points the path is close to the classical vacuum.
This part of the path integral is not semiclassical, but it
is known exactly, because it is governed by the vacuum
wavefunction. The two are glued at the matching points by
an ordinary Gaussian integral, which is stationary where
the momentum of the vacuum wavefunction equals the momentum
of the instanton.

Let $|0\rangle$ and $|2\pi\rangle$ be the vacuum states
localized in the well at $x=0$ and in its copy at $2\pi$
(they are not position eigenstates). On the circle these
are the same state, but for the winding-one amplitude it
is convenient to think of the potential along the line.
Their wavefunctions are $\psi_{\rm vac}(x)$ and
$\psi_{\rm vac}(x-2\pi)$, which equals $\psi_{\rm vac}
(2\pi-x)$ since $\psi_{\rm vac}$ is even. Up to corrections
of order $t$ they are eigenstates with energy $E_0$. In the
dilute gas, the part of the Euclidean amplitude in which
the path winds once is
\begin{equation}
\langle 2\pi|\,\e^{-HT}\,|0\rangle_{\rm one\ instanton}
   =T\,t\,\e^{-E_0T},
\label{eq:A1def}
\end{equation}
which defines $t$. Summing the dilute gas gives the band
\eqref{eq:band}. As a path integral, the left-hand side is
\begin{equation}
\langle 2\pi|\,\e^{-HT}\,|0\rangle
 =\int_{0}^{2\pi}\!\mathcal Dx\;\e^{-S[x]},\qquad
 S[x]=\int_0^T\!\dd\tau\,\big(\tfrac12\dot x^2+V\big),
\label{eq:PIdef}
\end{equation}
where the limits indicate that the paths start in the vacuum at
$x=0$ and end in the vacuum at $2\pi$; since $|0\rangle$ and
$|2\pi\rangle$ are not position eigenstates, the end points are
weighted by the vacuum wavefunctions. We evaluate it in five steps.

\begin{figure}[t]
\centering
\includegraphics[width=\textwidth]{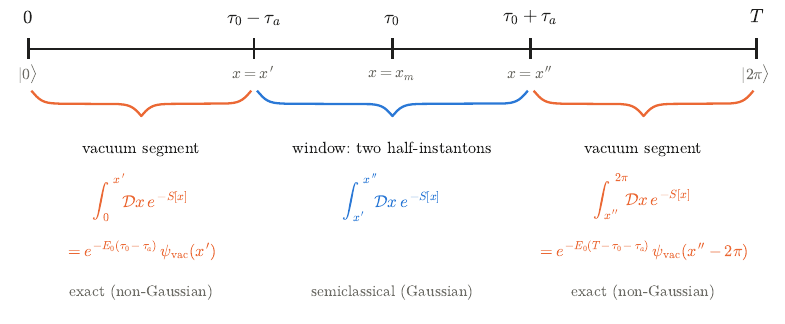}
\caption{The cut of the one-instanton path integral \eqref{eq:cut} in
Euclidean time. The path crosses the barrier top $x_m$ at $\tau_0$ and
passes the matching points $x'$ and $x''$ at $\tau_0\mp\tau_a$. The two
long vacuum segments are evaluated exactly and give the vacuum
wavefunction at the matching points (Step 2). The short window between
them is semiclassical: two half-instantons with fixed end points, joined
at the barrier top with the flux $\dot x_m$ (Step 3). The integrals over
$x'$ and $x''$ glue the pieces together (Step 4).}
\label{fig:cut}
\end{figure}

\paragraph{Step 1: cut the path.}
Let $\tau_0$ be the time at which the path crosses the top
of the barrier $x_m=\pi$, and $\tau_a$ the time the instanton
needs to go from $x_a$ to $x_m$, so that the instanton passes
the matching points at $\tau_0\mp\tau_a$. Cutting every path in
\eqref{eq:PIdef} at these two times, and integrating over the
positions $x'$ and $x''$ it has there, is the composition law of
the path integral, or equivalently the insertion of complete sets
of position states:
\begin{equation}
\langle 2\pi|\,\e^{-HT}\,|0\rangle
  = \int\dd x'\,\dd x''\;\langle 2\pi|\,
     \e^{-H(T-\tau_0-\tau_a)}\,|x''\rangle\,\langle x''|\,
     \e^{-2H\tau_a}\,|x'\rangle\,\langle x'|\,
     \e^{-H(\tau_0-\tau_a)}\,|0\rangle .
\label{eq:cut}
\end{equation}
Each factor is itself a path integral over the corresponding time
interval. The middle factor,
\begin{equation}
\mathcal K(x'',x';2\tau_a)\equiv\langle x''|\e^{-2H\tau_a}|x'\rangle
 =\int_{x(\tau_0-\tau_a)=x'}^{x(\tau_0+\tau_a)=x''}\!\mathcal Dx\;\e^{-S[x]},
\end{equation}
is the propagator across the window
$[\tau_0-\tau_a,\tau_0+\tau_a]$, which contains the instanton.
The outer factors are the vacuum segments, in which the instanton
tail is overwhelmed by the zero-point motion of the vacuum.
\Cref{fig:cut} summarizes this factorization.
The three path integrals are evaluated differently: the vacuum
segments exactly (Step 2), the window semiclassically (Step 3).

\paragraph{Step 2: the vacuum segments.}
Since $|0\rangle$ and $|2\pi\rangle$ are eigenstates up to $O(t)$,
\begin{align}
\langle x'|\e^{-H(\tau_0-\tau_a)}|0\rangle
 &=\int_{x(0)=0}^{x(\tau_0-\tau_a)=x'}\!\mathcal Dx\;\e^{-S[x]}
  = \e^{-E_0(\tau_0-\tau_a)}\,\psi_{\rm vac}(x'),\nonumber\\
\langle 2\pi|\e^{-H(T-\tau_0-\tau_a)}|x''\rangle
 &=\int_{x(\tau_0+\tau_a)=x''}^{x(T)=2\pi}\!\mathcal Dx\;\e^{-S[x]}
  = \e^{-E_0(T-\tau_0-\tau_a)}\,\psi_{\rm vac}(x''-2\pi).
\label{eq:vacseg}
\end{align}
The first is the sum over paths that start in the vacuum at $x=0$ and
reach the matching point $x'$ at time $\tau_0-\tau_a$; the second is
the vacuum of the well at $2\pi$ evaluated at a point near its own
minimum, $x''\simeq2\pi-x_a$. This is the vacuum part of the path
integral, evaluated exactly rather than semiclassically: the paths in
it stay near the bottom of the well, where there is no small parameter,
but their sum is fixed by the spectral decomposition, without any saddle
point. Since the vacuum is non-Gaussian we need $\psi_{\rm vac}$ and
$E_0$ as input (\cref{sec:vacuum}). For a harmonic well the same path
integral is Gaussian, and this step reproduces the usual determinant of
the vacuum. The total time $T$ is macroscopic, $T\gg\tau_q$, and the
crossing time $\tau_0$ is generically of order $T$, so both vacuum
segments are long compared with $\tau_q$, while the window $2\tau_a$ is
short. Had we started from position eigenstates at $x=0$ and $2\pi$
instead of the vacuum states, the long segments would project onto the
ground state, up to terms of order $\e^{-(E_1-E_0)(\tau_0-\tau_a)}$, with
the same result. The two segments together contribute
$\e^{-E_0(T-2\tau_a)}\,\psi_{\rm vac}(x')\,\psi_{\rm vac}(x''-2\pi)$,
independent of $\tau_0$.

\paragraph{Step 3: the window, two half-instantons and the
collective coordinate.}
Inside the window the path integral is semiclassical: the classical
path from $x'\simeq x_a$ to $x''\simeq2\pi-x_a$ in time $2\tau_a$
is the instanton, and the action is of order $S_0\gg\hbar$.
Its freedom to shift in time is the zero mode. We use the crossing
time $\tau_0$ as the collective coordinate. At fixed $\tau_0$ the
window splits into two half-instantons with \emph{fixed} endpoints,
from $x'$ to $x_m$ and from $x_m$ to $x''$, each of which has no
zero mode, so that its path integral is evaluated by steepest
descent with a finite Van Vleck factor,
\begin{equation}
\mathcal K_{\rm half}(x_m,x';\tau_a)
 =\int_{x(-\tau_a)=x'}^{x(0)=x_m}\!\mathcal Dx\;\e^{-S[x]}
 \simeq \frac{\e^{-S(x_m,x';\tau_a)}}{\sqrt{2\pi\Dh}},\qquad
\end{equation}
where 
\begin{equation}
\Dh=\dot x_a\dot x_m\,I_a,\qquad
I_a=\int_{-\tau_a}^{0}\frac{\dd\tau}{\dot x^2},
\end{equation}
is the half-window Gelfand--Yaglom determinant
\eqref{eq:Dhalf}. The measure for the collective coordinate is
$\dd\tau_0\,\dot x_m$: shifting the crossing time by
$\delta\tau_0$ moves the path at the top of the barrier by
$\delta x_m=\dot x_m\delta\tau_0$, and $\dot x_m$ is the
Jacobian between a cut at fixed time and a cut at fixed
position. This is the content of the identity \eqref{eq:flux}:
the zero-mode Jacobian $\sqrt{S_0/2\pi}$ and the lifted zero
mode in $\det'M_L$ combine into the flux $\dot x_m$ through
the barrier top, divided by the half-window determinants.
Hence
\begin{equation}
\mathcal K(x'',x';2\tau_a)  \;\to\;
  \int_0^T\dd\tau_0\;\dot x_m\;
   \mathcal K_{\rm half}(x'',x_m;\tau_a)\,
   \mathcal K_{\rm half}(x_m,x';\tau_a),
\label{eq:fluxdecomp}
\end{equation}
and the $\tau_0$ integral gives the factor $T$ of \eqref{eq:A1def}.
For the zero-energy path the half-instanton action is
\begin{equation}
S(x_m,x_a;\tau_a)=\int_{x_a}^{\pi}\sqrt{2V}\,\dd x
  =\frac{S_0}{2}-W(x_a),\qquad
W(x)=\kappa\,(x-\sin x).
\label{eq:Shalf}
\end{equation}

\paragraph{Step 4: gluing, i.e.\ momentum matching.}
What remains of the path integral is the integral over the
positions $x'$ and $x''$ at the cuts. On each side it is an
ordinary integral over the matching point,
\begin{equation}
\int\dd x'\,\psi_{\rm vac}(x')\,\e^{-S(x_m,x';\tau_a)}
 = \int\dd x'\,\e^{-\Phi(x')},\qquad
\Phi(x')\equiv-\ln\psi_{\rm vac}(x')+S(x_m,x';\tau_a).
\end{equation}
In the overlap region $\psi_{\rm vac}$ has the asymptotic form
$\e^{-W_0(x')}$ with $W_0(x')=\int_0^{x'}\sqrt{2V}$, up to the
slowly varying prefactor and the $E_0$ corrections (restored
below). For the classical path from $x'$ to $x_m$ in the fixed
time $\tau_a$, with Euclidean energy $E(x')$,
\begin{align}
S(x_m,x';\tau_a) &=  W_E(x')-E\,\tau_a,\qquad
W_E(x') = \int_{x'}^{x_m}p_E\,\dd x,\nonumber\\
p_E  &=\sqrt{2(V+E)},\qquad
\tau_a=\int_{x'}^{x_m}\frac{\dd x}{p_E}.
\end{align}
Differentiating, and using $\partial_EW_E=\tau_a$,
\begin{equation}
\Phi'(x')=p_0(x')-p_E(x').
\end{equation}
So the exponent is stationary where the momentum carried by
the vacuum wavefunction, $p_0=\sqrt{2V}$, equals the initial
momentum of the classical path. That is the zero-energy path,
$E=0$, which holds at $x'=x_a$ by construction, and indeed at
any $x_a$ in the overlap region. For the curvature, differentiate
once more at $E=0$: the explicit $x'$-dependence of $p_E$
cancels against $p_0'$, and only the change of the energy
survives,
\begin{equation}
\Phi'' = -\frac{\partial p_E}{\partial E}\,\frac{\dd E}{\dd x'}
  =  -\frac{1}{\dot x_a}\,\frac{\dd E}{\dd x'} .
\end{equation}
The rate at which the energy of the path changes with its starting
point, at fixed $\tau_a$, follows from differentiating the last
relation in the display above:
\begin{equation}
0 = \frac{\dd\tau_a}{\dd x'}
  =-\frac{1}{p(x')}-\frac{\dd E}{\dd x'}
       \int_{x'}^{x_m}\frac{\dd x}{p^3}
         \qquad\Longrightarrow\qquad
\frac{\dd E}{\dd x'} = -\frac{1}{\dot x_a I_a},
\end{equation}
since $\int\dd x/p^3=\int\dd\tau/\dot x^2=I_a$. Hence
$\Phi''=1/(\dot x_a^2\,I_a)$, and with $\Dh=\dot x_a
\dot x_mI_a$ we obtain
\begin{equation}
  \sqrt{\frac{2\pi}{\Phi''}}\times\frac{1}{\sqrt{2\pi\Dh}}=\sqrt{\frac{\dot x_a}{\dot x_m}} .
\label{eq:curv}
\end{equation}
The quantity $I_a=\int\dd\tau/\dot x^2$, which carries the whole
$L^5$ divergence of $\det'M_L$, cancels between the Van Vleck
determinant and the width of the endpoint integral. Each side
therefore contributes
\begin{equation}
\psi_{\rm vac}(x_a)\,\e^{W(x_a)}\,
  \sqrt{\frac{\dot x_a}{\dot x_m}}\;\e^{-S_0/2},
\label{eq:psim}
\end{equation}
where $\e^{W(x_a)}$ comes from the half-instanton action
\eqref{eq:Shalf} and $\psi_{\rm vac}(x_a)$ carries the prefactor
and the $E_0$-dependence of the vacuum wavefunction.

\paragraph{Step 5: assemble.}
Collecting Steps 2--4,
\begin{equation}
\langle 2\pi|\e^{-HT}|0\rangle_{\rm one\ instanton}
  =  T\,\e^{-E_0T}\;\dot x_m\,\Big[\psi_{\rm vac}(x_a)\,
    \e^{W(x_a)+E_0\tau_a}\sqrt{\frac{\dot x_a}{\dot x_m}}\,
    \e^{-S_0/2}\Big]^2,
\end{equation}
and comparing with \eqref{eq:A1def},
\begin{equation}
t=\dot x_a\,\psi_{\rm vac}(x_a)^2\,
  \e^{2W(x_a)+2E_0\tau_a}\;\e^{-S_0}.
\label{eq:tglued}
\end{equation}
This is independent of $x_a$ as long as $x_a$ is in the
overlap region: moving $x_a$ outward trades vacuum wavefunction
for instanton action, and the two are matched. The factor
$\e^{2E_0\tau_a}$ is the vacuum energy accumulated during
the time the instanton spends inside the window.

At leading order, in the overlap region, $\psi_{\rm vac}$ is
the quartic vacuum \eqref{eq:B}:
\begin{equation}
\psi_{\rm vac}(x)=\ell^{-1/2}\varphi_0(x/\ell)
 \simeq \frac{B\,G^{-1/12}}{x}
   \exp\!\Big(-\frac{\kq x^3}{3}-\frac{E_0}{\kq x}\Big).
\end{equation}
The instanton gives $\dot x_a\simeq\kq x_a^2$, $W(x_a)\simeq\kq
x_a^3/3$ and $\tau_a=\cot(x_a/2)/\kappa\simeq1/(\kq x_a)$. Every
$x_a$-dependence cancels in \eqref{eq:tglued}, including the $e_0$
term against $\e^{2E_0\tau_a}$, and the result is
\eqref{eq:tleading}: the path integral, with its vacuum segments
evaluated exactly, reproduces the amplitude of \cref{sec:herring}.
The two derivations are independent: the first never mentions an
instanton, the second never uses the wavefunction beyond the matching
point. Indeed the bracket in Step 5 is $\psi_L(x_m)\e^{E_0\tau_a}$,
the single-well state carried to the top of the barrier, so
\eqref{eq:tglued} is \eqref{eq:herring} with the flux through
the barrier playing the role of the collective-coordinate measure.

Compared with the determinant formula \eqref{eq:coleman}, the
instanton part is unchanged: the zero-mode Jacobian and $\det'M_L$
have been traded, through \eqref{eq:flux}, for the flux $\dot x_m$
and the half-window determinant $\Dh$, both finite. What changed is
the vacuum part. In \eqref{eq:coleman} the vacuum enters through a
Gaussian determinant $\det(-\partial_\tau^2+\omega^2)_L$; in
\eqref{eq:tglued} it enters through the wavefunction at the matching
point. The two formulas agree exactly if the Gaussian determinant is
replaced by an effective vacuum determinant built from
$\psi_{\rm vac}$,
\begin{equation}
\Dvac(L) = 2\pi\Big[\psi_{\rm vac}(x_a)^2\,
    \e^{2W(x_a)+2E_0\tau_a}\,\dot x_a^2\!
    \int_{-\tau_a}^0\frac{\dd\tau}{\dot x^2}\Big]^2
 = \frac{8\pi}{25}\,B^4\,G^{5/3}\,L^6\quad(\tau_a=L),
\label{eq:Dvac}
\end{equation}
which for harmonic minima reduces to $\det(-\partial_\tau^2+
\omega^2)_L=\e^{2\omega L}/2\omega$. The power $L^6$ cancels the
$L^6$ of $\det'M_L$ (in $\tau$ units) exactly as $\e^{2\omega L}$
does in the harmonic case; the free determinant $2L$ is too small
by $L^5$, which is the divergence found in \cref{sec:det}. The
four factors in \eqref{eq:Dvac} have a simple meaning: the vacuum
density at the point where the instanton tail sits at time $L$, the
removal of the tail suppression already counted in $\e^{-S_0}$, the
vacuum energy for the time spent inside the window, and the width of
the endpoint integral. For a Gaussian vacuum, $\psi_{\rm vac}=
(\omega/\pi)^{1/4}\e^{-\omega x^2/2}$, they combine into exactly
$\det(-\partial_\tau^2+\omega^2)_L$, which is why the determinant
formula and \eqref{eq:tglued} are the same formula for harmonic
minima \cite{Garg}. Numerically, at $S_0=200$, \eqref{eq:tglued}
with the numerically computed single-well state reproduces $t$ to
better than $1\%$ for any cut with $x_a\gtrsim3\ell$, while the
determinant formula with a free reference has no plateau and falls
as $L^{-5/2}$.

\subsection{Corrections: the first term of the one-instanton
series}
\label{sec:corrections}

The leading result \eqref{eq:tleading} used the pure quartic vacuum. 
The first correction is the first term of the perturbative expansion
around the instanton. In the path integral it would come from the 
$-\tfrac16\ell^2y^6$ term of the well \eqref{eq:Hwell} entering 
the vacuum segments and the gluing of Step 4 at one order beyond
Gaussian. It is much simpler to compute it from \eqref{eq:herring},
which is exact to all orders in this expansion: the single-well 
state is obtained by ordinary first-order perturbation theory 
around the quartic oscillator and carried to the top of the 
barrier. At higher orders this is a Bender--Wu-type recursion 
\cite{SulejmanpasicUnsal}. 

 Write the well as $y^4+\varepsilon y^6$ with $\varepsilon=
-\ell^2/6$, and let $\varphi_\varepsilon$ be its ground state, 
with energy $e_\varepsilon=e_0+\varepsilon e^{(1)}+\dots$, $e^{(1)}=
\langle y^6\rangle$. Its asymptotic form is the WKB form of 
the perturbed well, expanded to first order in $\varepsilon$:
\begin{equation}
\varphi_\varepsilon(y)
  \simeq \frac{B_\varepsilon}{y}
       \Big(1-\frac{\varepsilon y^2}{4}\Big)
       \exp\!\Big[-\frac{\sqrt2y^3}{3}
         -\frac{\sqrt2\,\varepsilon y^5}{10}
         -\frac{e_\varepsilon}{\sqrt2\,y}
         -\frac{e_0\,\varepsilon\,y}{2\sqrt2}\Big].
\label{eq:Beps}
\end{equation}
Inserting this into \eqref{eq:tglued} (equivalently, carrying 
it to the top of the barrier in \eqref{eq:herring}), every 
$\varepsilon$-dependent factor in the exponent and in the 
prefactor is canceled by the corresponding term of $W(x_a)$, 
$E_0\tau_a$ and $\dot x_a$, exactly as at leading order: the 
$\varepsilon y^5$ term is the $x^5$ term of $W$ in \eqref{eq:Wexp},
the $\varepsilon y$ term is the $x^6$ correction to $\tau_a$, 
and $(1-\varepsilon y^2/4)^2$ is canceled by $\dot x_a$. 
What survives is the normalization:
\begin{equation}
\frac{t}{t_{\rm lead}} = \Big(\frac{B_\varepsilon}{B}\Big)^2
 = 1+2\varepsilon\,\frac{\partial\ln B_\varepsilon}         
                        {\partial\varepsilon}\Big|_0
    +O(\varepsilon^2)
= 1 - \frac{\ell^2}{3}\,\frac{\partial\ln B_\varepsilon}
                             {\partial\varepsilon}\Big|_0 .
\end{equation}
The derivative is obtained from the first-order wavefunction 
$\varphi_1=\partial_\varepsilon\varphi_\varepsilon$, which 
satisfies $(h_0-e_0)\varphi_1=(e^{(1)}-y^6)\varphi_0$. The ratio
$u=\varphi_1/\varphi_0$ obeys the linear equation $u''+
2(\varphi_0'/\varphi_0)u'=2(y^6-e^{(1)})$, integrated inward from 
large $y$ with $\langle\varphi_0|\varphi_1\rangle=0$. Subtracting 
the growing terms of \eqref{eq:Beps} from $u$ and extrapolating in $1/y$ gives
\begin{equation}
\frac{\partial\ln B_\varepsilon}{\partial\varepsilon}\Big|_0
 = 0.41558,
 \end{equation}
which implies 
\begin{equation}
t=\sqrt2\,B^2\,G^{1/3}\,
   \e^{-S_0}\Big[1-0.1385\,G^{-1/3}+O(G^{-2/3})\Big].
\label{eq:tcorr}
\end{equation}
A fit to the numerical data up to $S_0=2500$ (Appendix~\ref{app:numerics})
gives $-0.138\pm0.001$ for the coefficient and reproduces the 
leading constant $\sqrt2B^2$ to $4\times10^{-5}$. The remaining
deviation is of the next order, $t/t_{\rm corr} - 1 \simeq 
0.0040\,G^{-2/3}$ over four orders of magnitude in $G$, with no sign 
of a term of order $\hbar/S_0\propto G^{-1/2}$. The same 
perturbation theory gives the $-0.00196\,G^{-1/3}$ term of 
the vacuum energy in \eqref{eq:E0series}. Written in terms 
of $S_0$, the relative correction to $t$ is $0.94\,(\hbar/
S_0)^{2/3}$, which decreases slowly.

The leading result \eqref{eq:tleading} and the correction in
\eqref{eq:tcorr} are the first two terms of the one-instanton 
series, which \eqref{eq:herring} generates to all orders from 
the all-orders single-well state. The difference
from harmonic minima is that the perturbative state is built 
around the quartic oscillator, so the series runs in powers of 
$\ell^2\propto\hbar^{2/3}$ rather than $\hbar$. Higher orders 
need the $y^8,y^{10},\dots$ terms of the well and the higher-order
WKB corrections under the barrier, which we have not pursued.

\subsection{Comparison with the numerical spectrum}
\label{sec:comparison}

\begin{figure}[t]
\centering
\includegraphics[width=\textwidth]{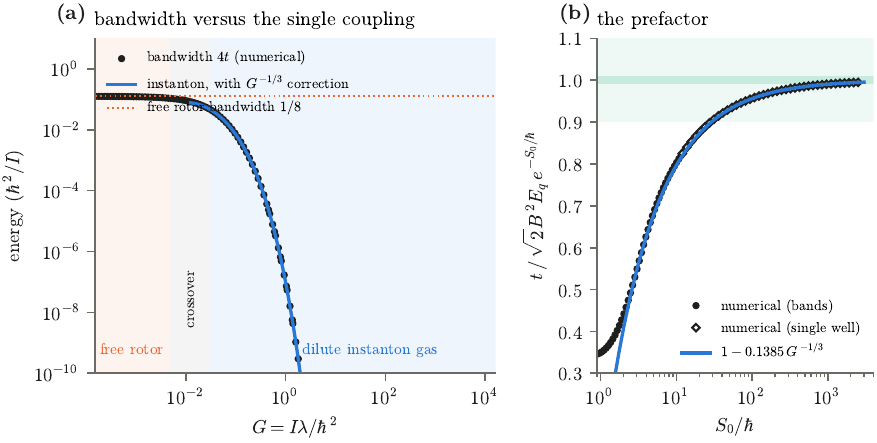}
\caption{(a) The bandwidth $4t$ versus $G$, computed numerically 
(dots), with the instanton prediction \eqref{eq:tcorr} and the 
free-rotor bandwidth $1/8$. (b) The numerical $t$ divided by the 
leading prediction $\sqrt2B^2E_q\e^{-S_0}$: the approach to $1$ 
is slow, but $1-0.1385\,G^{-1/3}$ tracks the data down to 
$S_0\approx3$.}
\label{fig:matching}
\end{figure}

The prediction for the lowest band is \eqref{eq:band} with $E_0$ 
from \eqref{eq:E0series} and $t$ from \eqref{eq:tcorr} ($\theta$ 
is the flux through the circle, \cref{sec:circle}). The problem 
is not exactly solvable, so we compare with a numerical solution 
of the Schr\"odinger equation, accurate to many more digits than
shown: band diagonalization in a Bloch--Fourier basis for 
$S_0\lesssim22$, and the amplitude formula \eqref{eq:herring} 
evaluated with the numerically computed single-well state for 
larger $S_0$ (Appendix~\ref{app:numerics}). \Cref{tab:matching} and 
\cref{fig:matching} show the result. The match is remarkable: 
with the first correction included, the instanton prediction 
is within $1\%$ of the numerical value already at $S_0\approx3$,
where the band is barely narrower than the free-rotor band.

\begin{table}[t]
\centering
\small
\begin{tabular}{@{}rrcccccc@{}}
\toprule
$G$ & $S_0$ & $E_0$ (num.) & $E_0$ from \eqref{eq:E0series} 
    & $t$ (num.) & $t/t_{\rm lead}$ & $t/t_{\rm corr}$\\
\midrule
0.031 & 3.14 & 0.15681 & 0.16006 & $1.140\times10^{-2}$ & 0.560 & 0.997\\
0.078 & 4.97 & 0.23437 & 0.23684 & $3.038\times10^{-3}$ & 0.682 & 1.008\\
0.313 & 9.93 & 0.40549 & 0.40625 & $3.943\times10^{-5}$ & 0.802 & 1.006\\
1.563 & 22.21 & 0.72903 & 0.72927 & $3.450\times10^{-10}$ & 0.883 & 1.002\\
15.63 & 70.25 & 1.62495 & 1.62500 & $1.096\times10^{-30}$ & 0.945 & 1.000\\
156.3 & 222.1 & 3.55327 & 3.55328 & $2.623\times10^{-96}$ & 0.974 & 1.000\\
\bottomrule
\end{tabular}
\caption{Numerical band data versus the instanton predictions (units 
$\hbar^2/I$). $t_{\rm lead}$ is \eqref{eq:tleading} and $t_{\rm corr}$
includes the factor $1-0.1385\,G^{-1/3}$ of \eqref{eq:tcorr}. The 
numerical values come from band diagonalization ($S_0\le22.2$) and 
from \eqref{eq:herring} with the numerically computed single-well 
state ($S_0\ge22.2$); the two methods agree to $10^{-5}$ where they 
overlap.}
\label{tab:matching}
\end{table}

Two features stand out. The exponent and the dilute-gas structure 
become reliable very early, at $S_0\approx3$. The \emph{prefactor}, 
by contrast, converges slowly: the leading formula is off by $23\%$ 
at $S_0=8$ and by $1\%$ only at $S_0\approx900$, because its 
corrections are powers of $(\tau_I/\tau_q)^2$, the size of a single 
instanton relative to the vacuum time, and that ratio decreases only 
as $G^{-1/3}$. Once the first correction is included the error is 
$O(G^{-2/3})$, below $1\%$ throughout.

\section{From the free rotor to the dilute instanton gas}
\label{sec:circle}

A flux $\theta$ through the circle, i.e.\ the boundary condition 
$\psi(x+2\pi)=\e^{i\theta}\psi(x)$, is equivalent to the Hamiltonian
\begin{equation}
H(\theta)=\frac12\Big(-i\partial_x+\frac{\theta}{2\pi}\Big)^2
  + 16G\sin^4\frac x2
\label{eq:Htheta}
\end{equation}
on periodic functions. The angle $\theta$ is the 
topological angle: an instanton winds once around the circle and 
picks up $\e^{i\theta}$, an anti-instanton $\e^{-i\theta}$. The 
ground-state energy $E_0(\theta)$ interpolates between two limits.

For the \emph{free rotor} ($G=0$) the levels are $E_k(\theta) = 
\tfrac12\big(k+\theta/2\pi\big)^2$, $k\in\mathbb Z$. The ground 
state is multi-branched, $E=\theta^2/8\pi^2$ for $|\theta|\le\pi$, 
with a level crossing at $\theta=\pi$, and all bands touch. For 
small $G$, $E_0(\theta)=\theta^2/8\pi^2+6G+O(G^2)$, and the crossing 
at $\theta=\pi$ is avoided with a gap $8G+O(G^2)$, from the $\cos x$
Fourier component of $16\sin^4(x/2)=6-8\cos x+2\cos2x$. In the 
\emph{dilute instanton gas} ($S_0\gg1$), summing independent, 
Poisson-distributed events \cite{Coleman,Shapiro},
\begin{equation}
Z(\theta,T) \simeq \e^{-E_0T}\sum_{n,\bar n}
   \frac{(tT)^{n+\bar n}}{n!\,\bar n!}\,
    \e^{i\theta(n-\bar n)}
 = \exp\big[-E_0T+2tT\cos\theta\big],
\end{equation}
which reproduces the band \eqref{eq:band}: parabolas become a cosine.

\begin{figure}[t]
\centering
\includegraphics[width=\textwidth]{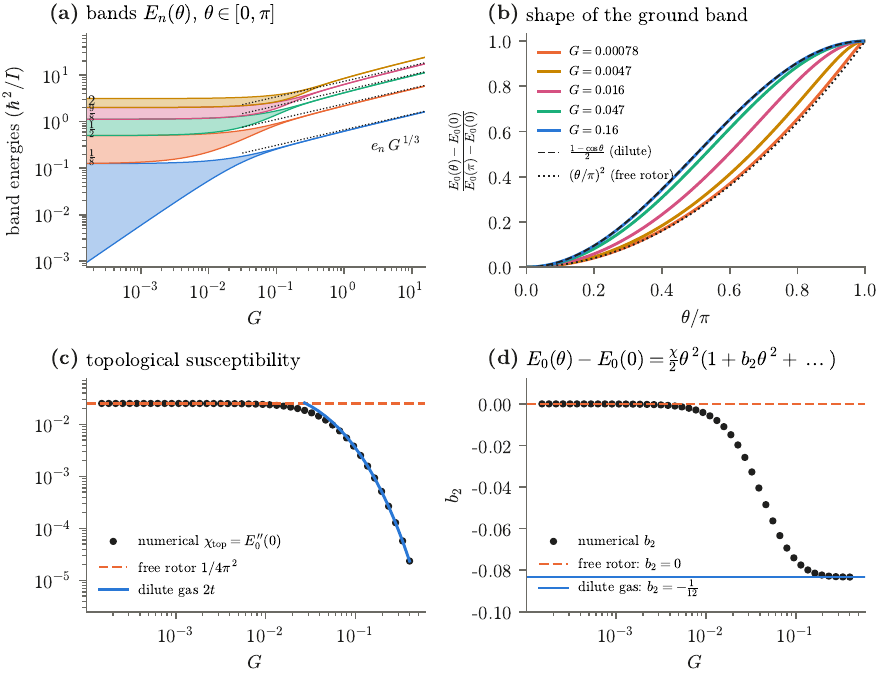}
\caption{From the free rotor to the dilute instanton gas. (a) The 
five lowest bands $E_n(\theta)$, $\theta\in[0,\pi]$, versus $G$. 
At $G\to0$ they touch at the free-rotor values $\frac18,\frac12,
\frac98,2$. At large $G$ they shrink onto the quartic-oscillator 
levels $e_nG^{1/3}$ (dotted). (b) Normalized shape of the ground 
band, from a parabola to a cosine. (c) Topological susceptibility. 
(d) The quartic coefficient $b_2$, from $0$ to $-1/12$.}
\label{fig:theta}
\end{figure}

\Cref{fig:theta} shows how the spectrum interpolates between these 
limits, through three simple diagnostics. The topological 
susceptibility $\chi_{\rm top}\equiv E_0''(0)$ goes from
$1/4\pi^2=0.0253$ (free rotor) to $2t$ (dilute gas). The shape
coefficient $b_2$, defined by 
\begin{equation}
E_0(\theta)-E_0(0)=\tfrac{\chi}{2}\theta^2
  \big(1+b_2\theta^2+\dots\big),
\end{equation}
goes from $0$ (a parabola) to $-1/12$ (a cosine). The ratio
\begin{equation}
R=\frac{E_0(\pi)-E_0(0)}
        {\pi^2\chi/2}
\end{equation}
goes from $1$ to $4/\pi^2=0.405$. The crossover is centered at
$G \approx 0.04$--$0.05$, i.e.\ $S_0\approx 3.5$, where $\chi$ 
and $b_2$ are halfway between their limits. At $S_0=6.1$ ($G=0.116$),
for example, $\chi_{\rm top} = 2.42\times10^{-3}$ against $2t = 
2.45\times10^{-3}$ from \eqref{eq:tcorr}. In quantum mechanics 
it is smooth: there is no phase transition.

 As $G$ decreases through $G\approx0.016$--$0.05$ ($S_0\approx2$--$4$),
the instanton picture melts: the exponential dependence of $t$ on 
$\sqrt\lambda$ is lost, the band shape turns from a cosine into 
free-rotor parabolas, and $\chi_{\rm top}$ saturates at $1/4\pi^2$. 
The classical instanton survives at every coupling (it has no 
negative mode). What melts is the semiclassical approximation
once $S_0$ becomes of order $\hbar$.

 There is an analogue of the non-degeneracy theorem for the circle, 
and it fixes which instanton effects can and cannot cancel. At 
$\theta=0$ the Euclidean evolution operator $\e^{-HT}$ has a 
strictly positive kernel, being a path integral with a positive 
weight, so by the Perron--Frobenius theorem the ground state is 
unique and nodeless. This holds on any connected configuration 
space \cite{ReedSimonIV}. For a periodic potential on the line, 
or equivalently on the circle with flux $\theta$, more is true: 
the band functions of a one-dimensional periodic Schr\"odinger 
operator are strictly monotone in the quasi-momentum between 
$0$ and $\pi$ \cite{ReedSimonIV}. Hence $E_0(\pi)>E_0(0)$ for 
every $G$, the band can never be flat, and the instanton amplitude 
$t$ can never vanish. What can fail is positivity. For $\theta\ne0$ 
the weight $\e^{i\theta Q}$ is complex, and if the potential has 
an additional symmetry, such as the shift $x\to x+\pi$ of a potential
with two equivalent minima per period, a mixed anomaly between 
that symmetry and the $\theta$-periodicity forces an exact two-fold 
degeneracy at $\theta=\pi$ \cite{GKKS}: there the contributions of 
the two kinds of tunneling events cancel. The potential 
$16 \lambda \sin^4(x/2)$ has a single minimum per period and 
no such symmetry, and the gap at $\theta=\pi$ stays open
(\cref{fig:theta}a).

\section{The quartic double well}
\label{sec:dw}

Nothing in the analysis above used the periodicity of the potential 
in an essential way. To see what is universal and what is not, 
consider the quartic double well \eqref{eq:dw},
\begin{equation}
H = -\frac12\partial_x^2+\lambda\,x^4(1-x)^4 ,
\label{eq:Hdw}
\end{equation}
in units $\hbar=m=1$, with the minima at $x=0$ and $x=1$ as the unit 
of length. Both minima are quartic with the same coefficient,
$V\simeq\lambda x^4$ and $V\simeq\lambda(1-x)^4$, and the barrier, 
of height $\lambda/256$, sits at $x=\frac12$. The coupling $\lambda$
plays the role of $G$. This model was considered by Patrascioiu 
\cite{Patrascioiu} and by Mottola and Rouet \cite{MottolaRouet}, 
who concluded that the dilute instanton gas description fails. Below 
we show that it actually works, because the vacuum generates a gap
dynamically: once the instanton is joined to the exact quartic vacuum, 
the quantum interactions decay exponentially beyond $\tau_q=
\lambda^{-1/3}$, at the rate set by the gap of each well, $(e_1-e_0)
E_q=1.726\,E_q$, and the amplitude agrees with the numerical splitting.

The classical side is again elementary. The instanton solves 
$\dot x=\sqrt{2\lambda}\,x^2(1-x)^2$, whose solution is known in
implicit closed form,
\begin{equation}
\sqrt{2\lambda}\,(\tau-\tau_0) =
  \frac{1}{1-x}-\frac1x+2\ln\frac{x}{1-x},
\label{eq:dwinst}
\end{equation}
with action $S_0=\int_0^1\sqrt{2V}\,\dd x=\sqrt{2\lambda}/30$ and 
power-law tails $x\simeq1/(\sqrt{2\lambda}\,|\tau|)$: it is again an algebraic instanton. Its zero mode 
$\dot x\propto x^2(1-x)^2$ decays as $1/\tau^2$, so it is again a 
bound state at the threshold of a gapless continuum, the determinant
formula with a free reference is infrared divergent in the same way,
and linear superposition fails for the same reason. The $I\Ib$ interaction 
is controlled by the hovering near the quartic hilltop of $-V$, and 
is therefore the same universal law \eqref{eq:interactions}, 
$S_{I\Ib}\simeq-3.94\,\hbar(\tau_q/T)^3$ with $\tau_q=\lambda^{-1/3}$. 
The only change is that the regular time offset $T_0$ of 
\eqref{eq:Tsep4} is replaced by a logarithm, a relative correction 
of order $(\tau_I/T)\ln(T/\tau_I)$. There is no $II$ channel: an
instanton must be followed by an anti-instanton.

The quantum side is where the two potentials differ. Near each 
minimum the vacuum is the same quartic oscillator, with 
$\ell=\lambda^{-1/6}$ and $E_q=\lambda^{1/3}$, and the matched
calculation of \cref{sec:matching} goes through unchanged. The
splitting of the two lowest levels is $\Delta E=2t$, and at 
leading order
\begin{equation}
t = \sqrt2\,B^2\,\lambda^{1/3}\,\e^{-S_0},\qquad 
S_0=\frac{\sqrt{2\lambda}}{30},
\label{eq:dwlead}
\end{equation}
with the same constant $\sqrt2B^2=1.4953$ as \eqref{eq:tleading}. 
The leading amplitude is universal: it depends only on the quartic
coefficient at the minimum and on $S_0$. That $t$ cannot vanish is
guaranteed in advance: in one dimension the bound states of a 
confining potential are non-degenerate \cite{LL}, so the two lowest
levels must split and the one-instanton amplitude must be non-zero,
whatever the shape of the minima.

The corrections are not universal. In units of $\ell$ the well is
\begin{equation}
H = E_q\Big[-\tfrac12\partial_y^2+y^4-4\,\ell\,y^5
   +6\,\ell^2y^6-\dots\Big],
\end{equation}
which, unlike \eqref{eq:Hwell}, is not symmetric about its minimum: 
each well is softer on the side of the barrier. The first correction 
is therefore of order $\ell\propto\hbar^{1/3}$ rather than $\ell^2$.
Moreover, the time the instanton needs to travel from the matching
point to the barrier top acquires a logarithm, $\sqrt{2\lambda}\,
\tau_a=1/x_a+2\ln(1/x_a)-1+O(x_a)$, from the same asymmetry. In
\eqref{eq:tglued} the factor $\e^{2E_0\tau_a}$ then contains
$2\sqrt2e_0\,\ell\ln(1/x_a)$. The vacuum wavefunction of the perturbed 
well supplies $2\sqrt2e_0\,\ell\ln(x_a/\ell)$, the $x_a$-dependence 
cancels as it must, and a term $2\sqrt2e_0\,\ell\ln(1/\ell)$ is left 
over. Since this term comes from the factor $\e^{2E_0\tau_a}$, it is
natural to keep it in the exponent. Together with the first-order 
change of the tail constant $B$, computed as in \cref{sec:corrections} 
with the $y^5$ perturbation, this gives
\begin{equation}
t = \sqrt2\,B^2\,\lambda^{1/3}\,\e^{-S_0}
  \exp\!\Big[2\sqrt2\,e_0\,\ell\ln\frac1\ell-0.20\,\ell
     +O(\ell^2)\Big],\qquad \ell
 = \lambda^{-1/6}\simeq0.36\,(\hbar/S_0)^{1/3}.
\label{eq:dwcorr}
\end{equation}
This agrees with the numerically computed splitting: a fit for 
$200\le S_0\le10^6$ gives $1.89$ for the coefficient of 
$\ell\ln(1/\ell)$, against $2\sqrt2e_0=1.889$, and $-0.2$ for 
the coefficient of $\ell$, and with both terms included
\eqref{eq:dwcorr} is within $1\%$ of the numerical value from
$S_0\approx1000$ on, the remainder being of order $\ell^2$. The 
leading formula alone, by contrast, is still $23\%$ off at 
$S_0=1000$, whereas for $\sin^4(x/2)$ it is $1\%$ off at 
$S_0\approx900$. The lesson is general. For a non-Gaussian 
vacuum the leading one-instanton amplitude is fixed by the vacuum
constant $B$ and the instanton action alone, but the approach to 
it is controlled by fractional powers of $\hbar$, and how fractional 
depends on how the potential deviates from its leading term near the
minimum. In practice the corrected formula, not the leading one, is
what should be compared with the spectrum. 

\section{Relation to Yang--Mills instantons}
\label{sec:YM}

The feature that motivates this work is shared with instantons of 
Yang--Mills theory on $\mathbb R^4$ and of sigma models on $\mathbb 
R^2$: the vacuum the instanton lives in is non-Gaussian, the gap 
is generated dynamically, and the classical interaction between
instantons is a power law. In both theories the instanton is not, 
in general, weakly coupled. In Yang--Mills theory the coupling is 
set by its size, $g^2(1/\rho)$, and grows for large $\rho$, so 
only small instantons are under semiclassical control. The instanton
studied in this work has a fixed size and a small coupling 
$\hbar_{\rm eff}=\pi/S_0$. 

  Several elements of the analysis in this work have recognizable
counterparts in Yang--Mills theory. The BPST instanton has power-law 
tails, there are normalizable zero modes at the edge of a 
massless continuum \cite{tHooft}, and in a dilute-gas treatment 
on $\mathbb R^4$, with fluctuations normalized to the perturbative
vacuum, the integral over instanton sizes grows in the infrared 
\cite{tHooft,CDG}. In this sense instantons in theories with massless fields are algebraic,
a consequence of the scale invariance of the massless tail equation. In
quantum mechanics algebraic instantons are just as generic, since any
minimum with $V''=0$ produces them, but they have largely been overlooked;
the present model is their simplest realization.
In the quantum mechanical system the classical 
$I\Ib$ interaction is a power law at intermediate separation and 
is screened by the dynamically generated gap at large distance. 
The $\theta$-dependence interpolates between the free rotor ($b_2=0$) 
and the dilute gas ($b_2=-1/12$), the same coefficient that 
lattice studies use to identify dilute-gas behavior at high 
temperature \cite{VP}.

The sum over states of Appendix~\ref{app:spectral} has a direct field-theory 
counterpart, with the intermediate well replaced by the vacuum: the
propagator between the two cores becomes the vacuum propagator between
the sources set by the two instanton tails, and the sum over states
of the well becomes a sum over multi-particle states, which for a 
free light field reduces to an integral over momentum modes, 
\begin{equation}
\e^{-S_{\rm int}}\simeq\exp\big[\int\!
  \frac{\dd^dk}{(2\pi)^d}
   \,J_1(k)J_2(-k)\,\e^{ik\cdot R}/(k^2+m^2)\big].
\end{equation}
Each mode is then in a coherent state with Poisson weights, exactly 
as in the harmonic case of Appendix~\ref{app:harmonic}, and the interaction 
is the exchange of quanta. The Polyakov model is the clearest example 
\cite{Polyakov}: monopole-instantons interact through the classically 
massless dual photon, by a Coulomb law at short distances, and beyond
the inverse of the dual-photon mass $m_\gamma\propto\e^{-S_0/2}$, which
the monopoles themselves generate, the interaction is screened, as for
$T\gg\tau_q$ in our model; there the screened propagator is controlled 
because the dual photon is weakly coupled. In four dimensions the same 
picture, with a massless gluon and $m=0$, gives the dipole interaction 
of instantons through gluon exchange at separations $\rho\ll R\ll1/
\Lambda$ \cite{CDG}; at larger separations the gluon is not a weakly 
coupled field and there is no comparable controlled statement. The 
difference with our model is that in these theories the power law 
comes from the infrared end of a continuum of nearly free quanta,
whereas here the spectrum of the well is discrete and gapped, the 
tail equation is nonlinear at all distances, and the classical 
$T^{-3}$ law comes from a band of highly excited states.

We stress that the analogy is structural, and that it does not imply
that the same mechanism is at work in gauge theories. Here the
instanton has a fixed size, the infrared problem sits in its tails, 
and it is resolved by replacing the Gaussian vacuum reference by the true
vacuum wavefunction, which is known. In gauge theories and sigma models
there are other, controlled ways in which instantons and their
constituents become meaningful: in suitably compactified theories on 
$\mathbb R^3\times S^1$ and $\mathbb R\times S^1$ the dynamics is
weakly coupled and semiclassics is under control \cite{Unsal2008,
UnsalYaffe2008,ShifmanUnsal2008,ArgyresUnsal,DunneUnsalCPN,
DunneUnsalARNPS,Unsal2021,Poppitz2022,TanizakiUnsal2022,NguyenUnsal},
and in self-dual theories the sum over all instanton sectors can be 
carried out without the dilute-gas assumption \cite{HayashiUnsal2026a,
HayashiUnsal2026b}. The present model does not bear on these directly. 
What it isolates, in the simplest setting in which every step can be 
checked, is how a classically gapless vacuum enters an instanton 
amplitude: through its wavefunction rather than through a Gaussian 
determinant. It also differs from Yang--Mills theory in obvious ways: 
there is no size modulus, the crossover in $\theta$-dependence is 
smooth, two instantons repel classically, and there are no fermions 
and no index theorem.

\section{Conclusions and prospects}
\label{sec:conclusions}

We have carried out the instanton calculation for an algebraic 
instanton, that of a particle on a 
circle with a quartic minimum, from the classical solution to the 
comparison with the spectrum. The lesson is simple to state. With an 
anharmonic minimum the classical vacuum is gapless and non-Gaussian, 
and every part of the textbook construction that relied on the gap 
fails: the one-loop determinant is infrared divergent, superposition 
of instantons fails, and the interactions are power laws. All of it 
is repaired by one change: the vacuum enters the instanton amplitude
through its wavefunction, not through a Gaussian determinant. The
leading amplitude is then universal, $t=\sqrt2B^2G^{1/3}\e^{-S_0}$,
with a prefactor $\propto\hbar^{4/3}$; the corrections, in fractional
powers of $\hbar$, are not, and they depend on how the potential
deviates from $x^4$ near the minimum. The classical $T^{-3}$ 
interaction is fixed by the singularity of the period at the separatrix 
and is screened beyond the quantum time $\tau_q$ by the dynamically 
generated gap.

Two directions seem natural for further work. First, the resurgent 
structure, which this paper does not address and which remains 
open: the perturbative series around the vacuum runs in powers of 
$\hbar^{2/3}$, the $I\Ib$ interaction is a power law up to $\tau_q$,
and it is not clear how the cancellations between the Borel ambiguity
of perturbation theory and the $I\Ib$ amplitude \cite{DunneUnsal,
MarinoBook} are organized in such a theory. The amplitude formula of
\cref{sec:herring}, fed with the all-orders single-well state, is the
natural tool for generating the one-instanton series to high order.
Second, the same questions can be asked in multi-dimensional quantum
mechanics with degenerate minima, where the zero-point motion of a 
non-Gaussian vacuum must again be matched to the instanton tails, 
and where the asymmetry of the wells, as \cref{sec:dw} shows, controls
how fast the instanton result is approached.

\acknowledgments
 We thank Gerald Dunne for discussions. This work made 
extensive use of AI tools Claude Opus 5.5 (Anthropic) and 
ChatGPT-6 (OpenAI). The initial formulation of the problem as 
well as the initial analytic work was not based on AI. AI tools were 
used for some of the analytic derivations, numerical checks, 
preparation of figures, and initial drafts of the text. The 
authors checked all the analytic work, and rewrote the text. 
Numerical tests and cross checks are summarized in 
Appendix~\ref{app:numerics}. The authors take full responsibility for 
the work. H.~G.~is supported by the Gravity, Spacetime, and 
Particle Physics (GRASP) Initiative from Harvard University. 
R.~A., T.~S. and M.~\"U.~are supported by the U.S. Department 
of Energy, Office of Science, Office of Nuclear Physics under 
Award Number DE-FG02-03ER41260. M.~\"U.~is also supported by 
the Simons Foundation Grant ``Simons Collaboration on Confinement 
and QCD Strings''.

\appendix
\section{Instanton interactions as a sum over states}
\label{app:spectral}

The classical calculation of \cref{sec:separatrix} never refers to 
the spectrum of the well. Yet the decay rate of the harmonic
interaction, $\omega$, is exactly the gap $E_1-E_0$, and the 
quartic interaction is cut off at $T\sim\tau_q$ by the quantum 
gap. This appendix makes the connection explicit: the interaction 
is the excited-state part of a transfer matrix. For harmonic minima 
the sum over states exponentiates into the classical formula. For
quartic minima its saddle point is the separatrix result. In this 
appendix $\hbar$ is kept explicit.

\subsection{The interaction as the excited-state part of a 
transfer matrix}
\label{app:transfer}

\begin{figure}[t]
\centering
\includegraphics[width=\textwidth]{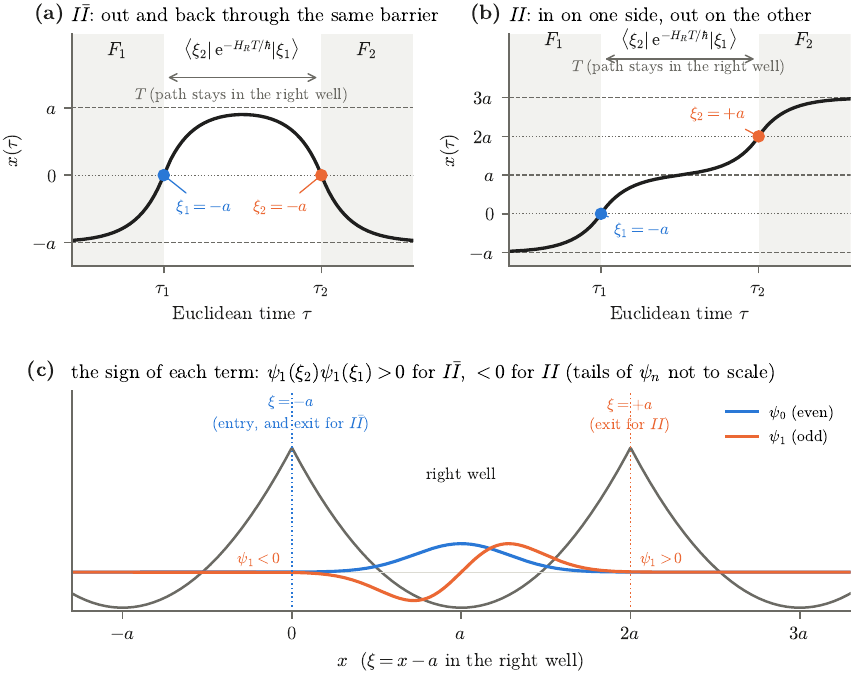}
\caption{What \eqref{eq:transfer} means. (a) An $I\Ib$ history: 
the instanton brings the particle from the well at $-a$ into the 
right well at $+a$; the anti-instanton takes it back a time $T$ 
later. (b) An $II$ history: the second instanton takes the particle 
on to the well at $3a$. The history is cut at $\tau_1$ and $\tau_2$,
where the path crosses a barrier top. The gray pieces, which contain
the events, give the $T$-independent factors $F_{1,2}$; the white 
piece is the propagator of the right well, with $\xi=x-a$. The curves
are the classical paths of the piecewise-harmonic model of 
Appendix~\ref{app:harmonic}. (c) The right well with $\psi_0$ (even) and 
$\psi_1$ (odd). For $I\Ib$ both cut points lie at $\xi=-a$, so 
$\psi_1(\xi_2)\psi_1(\xi_1)>0$; for $II$ they lie on opposite sides, 
so $\psi_1(\xi_2)\psi_1(\xi_1)<0$. The same holds for every odd $n$.}
\label{fig:transfer}
\end{figure}

\Cref{fig:transfer} shows the two histories. An instanton carries 
the particle into the right well, and a time $T$ later either an 
anti-instanton carries it back ($I\Ib$) or a second instanton 
carries it on into the next well ($II$). Cut each history at the 
moment $\tau_1$ the first event is over and the moment $\tau_2 
= \tau_1+T$ the second one starts, and measure positions from the
bottom of the right well, $\xi=x-a$. The two outer pieces, which
contain the events, are $T$-independent factors $F_{1,2}$. The 
middle piece stays in the right well, and its path integral is 
by definition the propagator of the well, $\langle\xi_2|
\e^{-H_RT/\hbar}|\xi_1\rangle$. If the cuts are placed where the 
path crosses the barrier tops, the cut points are fixed, $\xi_1=
\xi_2=-a$ for $I\Ib$ and $\xi_1=-a$, $\xi_2=+a$ for $II$, and 
the amplitude is
\begin{equation}
\mathcal A(T) = F_1\,F_2\;\langle\xi_2|\,
    \e^{-H_RT/\hbar}\,|\xi_1\rangle
 = F_1F_2\sum_{n\ge0}\psi_n(\xi_2)\,\psi_n(\xi_1)\,
     \e^{-E_nT/\hbar}.
\label{eq:transfer}
\end{equation}
The $n=0$ term alone is the dilute gas, two independent events 
with the vacuum propagating in between. Everything else is the
interaction:
\begin{equation}
\e^{-S_{\rm int}(T)/\hbar}
  =  \sum_{n\ge0}R_n\,\e^{-(E_n-E_0)T/\hbar},  \qquad
R_n = \frac{\psi_n(\xi_2)\,\psi_n(\xi_1)}
           {\psi_0(\xi_2)\,\psi_0(\xi_1)} .
\label{eq:spectralA}
\end{equation}
This holds for any well. If the cut points are not fixed, 
$\psi_n(\xi_{1,2})$ is replaced by the overlap with $F_{1,2}$.
Two consequences are immediate. The \emph{range}: at large $T$ 
the sum is dominated by $n=1$ and decays at the rate $(E_1-E_0)/
\hbar$, the gap. The \emph{sign}: in a symmetric well $\psi_n$ 
has parity $(-1)^n$, so $R_n$ carries $[\operatorname{sign}
(\xi_1\xi_2)]^n$. For $I\Ib$ the path leaves through the barrier 
it came in by, all $R_n>0$, and the interaction is attractive. 
For $II$ it leaves through the opposite barrier, the odd terms 
are negative, and the interaction is repulsive (\cref{fig:transfer}c).

The large weights have a simple origin. Under the barrier the WKB
wavefunction of energy $E$ is $\psi_E \propto \exp[ \frac1\hbar 
\int^\xi\sqrt{2(V-E)}]$, so $\partial_E\ln\psi_E(\xi)=\tau(\xi)/
\hbar$, where $\tau(\xi)$ is the Euclidean time the zero-energy
trajectory needs to travel from the quantum region of the well to 
$\xi$. Hence
\begin{equation}
 R_n\,\e^{-(E_n-E_0)T/\hbar}
   \simeq \mathcal N_n^2\, 
   \e^{-(E_n-E_0)\,[T-\tau(\xi_1)-\tau(\xi_2)]/\hbar}
\label{eq:Rtime}
\end{equation}
and the spectral sum measures time not from the cores but from the 
moments the tails enter the quantum region.

\subsection{The harmonic case: exchange of quanta}
\label{app:harmonic}

Take a potential that is exactly harmonic in each well, $V = \tfrac12 
\omega^2(|x|-a)^2$, with instanton $x_I=a(1-\e^{-\omega\tau})$ for 
$\tau>0$ and $S_0=\omega a^2$. The cores are where the path crosses
$x=0$, i.e.\ $\xi=-a$, and between them the dynamics is exactly that 
of the oscillator centered at $a$.

The oscillator action from $\xi_1$ to $\xi_2$ in time $T$, minus the
two isolated tails, gives with $w=\e^{-\omega T}$
\begin{equation}
S^{I\Ib}_{\rm int}=-\,2S_0\,\frac{w}{1+w},\qquad
S^{II}_{\rm int}=+\,2S_0\,\frac{w}{1-w},
\label{eq:harmcl}
\end{equation}
which for $\omega T\gg1$ reduce to the familiar $\mp2S_0\,
\e^{-\omega T}$.

The same answer follows from the spectrum. With $\psi_n/\psi_0=H_n(y)/
\sqrt{2^nn!}$, $y=\xi\sqrt{\omega/\hbar}$, the cores sit far out in 
the classical tail, $y_1y_2=\pm S_0/\hbar$. For $n\ll S_0/\hbar$, 
$H_n(y)\simeq(2y)^n$, so the weights are Poisson, $R_n \simeq 
(2y_1y_2)^n/n!$, and since the levels are equally spaced the sum
exponentiates:
\begin{equation}
\e^{-S_{\rm int}/\hbar} \simeq 
   \sum_{n\ge0}\frac{\big(2y_1y_2\,\e^{-\omega T}\big)^n}{n!}
 = \exp\!\big(2y_1y_2\,\e^{-\omega T}\big)
    \quad\Longrightarrow\quad 
S_{\rm int}\simeq\mp\,2S_0\,\e^{-\omega T}.
\end{equation}
This is the classical result, obtained without any classical solution.
Mehler's formula does the sum exactly:
\begin{equation}
S_{\rm int}=\mp\,2S_0\,\frac{w}{1\pm w}
  \; + \;\frac{\hbar}{2}\ln\big(1-w^2\big).
\label{eq:harmexact}
\end{equation}
The first term is exactly \eqref{eq:harmcl}; the second is the 
one-loop correction.

Two points are worth mentioning. The $n$-th term is the exchange 
of $n$ quanta, each costing $\e^{-\omega T}$, so the rate $\omega$ 
of the classical formula is the gap. And the prefactor $2S_0/
\hbar\sim1/g$ is an occupation number, the mean $\bar n=(2S_0/
\hbar)\e^{-\omega T}$ of the Poisson distribution, so that 
$S_{\rm int}=\mp\hbar\bar n$: the classical tail is a coherent 
state carrying many quanta. Because a Poisson distribution over 
equally spaced levels exponentiates exactly, the many-quanta
(classical) and single-quantum regimes are described by one 
formula, which is why superposition gives the complete answer 
for harmonic minima.

\subsection{The quartic case: fugacity 
\texorpdfstring{$\e^{\pi}$}{e\^{}pi} per quantum}
\label{app:quartic}

For quartic minima \eqref{eq:spectralA} still holds, but neither
harmonic fact survives: the levels are not equally spaced, and the
weights are not Poisson. The sum nevertheless has a simple structure.

\begin{figure}[t]
\centering
\includegraphics[width=\textwidth]{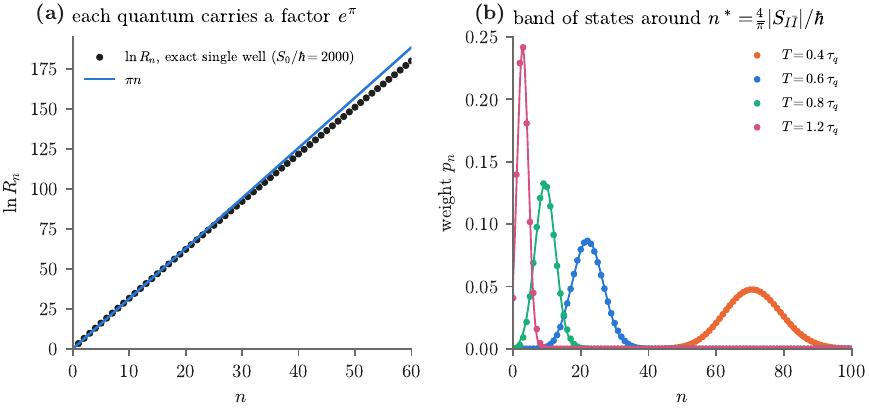}
\caption{The quartic interaction as a sum over states, for the 
numerically computed single-well states of $\sin^4(x/2)$ at $S_0/
\hbar=2000$, with the cuts at the instanton cores. (a) $\ln R_n$ 
against $\pi n$: each quantum carries a factor close to $\e^{\pi}$ 
(the slow drift at large $n$ comes from the deviation of $16 \sin^4 
(x/2)$ from $x^4$). (b) The weights $p_n\propto R_n\,\e^{-(E_n-E_0)T}$ 
in the $I\Ib$ channel (dots), against Gaussians centred at $n^* = 
\frac4\pi |S_{I\Ib}|/\hbar$ with width $\sqrt{3n^*/\pi}$ (lines).}
\label{fig:sumstates}
\end{figure}

Integrating $\partial_\varepsilon\ln\psi_\varepsilon=\tau(\xi;
\varepsilon)/\hbar$ from $0$ to $\varepsilon$, the singular 
(hovering) part of the time, $c/(2\kq\eta_t)$ with
$G\eta_t^4=\varepsilon$, gives exactly a quarter of the 
Bohr--Sommerfeld action of the well, $J(\varepsilon)=\oint p\,\dd x$. 
With two cores and $J(\varepsilon_n)=2\pi\hbar(n+\frac12)$,
\begin{equation}
R_n\simeq\e^{\pi n}
\label{eq:Rquartic}
\end{equation}
up to slowly varying prefactors (for the true cores of $\sin^4(x/2)$, 
where $T_0=0$). The enhancement $\e^{\pi}$ per quantum does not depend 
on $G$, on $\hbar$, or on where the core sits. The reason is that for 
$V=Gx^4$ the substitution $x\to ix$ maps the Euclidean motion in $-V$ 
onto real-time motion in $V$, so the hovering time near the top at 
energy $-\varepsilon$ is exactly half the real period at $\varepsilon$, 
and the hovering costs half the Bohr--Sommerfeld action, $\pi(n+
\frac12)$. For a harmonic $x^2$ the substitution flips the sign of the 
potential, and no such identity exists. \Cref{fig:sumstates}a confirms 
\eqref{eq:Rquartic} for the numerically computed single-well states of $
\sin^4(x/2)$ at $S_0/\hbar=2000$.

In the $I\Ib$ channel all terms are positive, and with $\varepsilon_n
\propto n^{4/3}$ the sum is dominated by its largest terms, $\sum_n 
\exp[\pi n-(\varepsilon_n-\varepsilon_0)T/\hbar]$. The stationarity 
condition, $\pi=(T/\hbar)\,\dd\varepsilon/\dd n$, is exactly the 
classical condition $T(E)=T$ with $E=-\varepsilon_{n^*}$: the dominant 
eigenstate is the one whose turning point is the turning point of the 
Euclidean orbit. Evaluating the exponent there gives $S_{I\Ib}=-
\frac\pi4\hbar n^*=-c^4/(12GT^3)$, which is \eqref{eq:interactions0}: 
the saddle point of the sum over states is the Legendre transform 
\eqref{eq:legendre}, and the separatrix derivation is its classical 
limit. The states that matter form a Gaussian band around 
$n^*=\frac4\pi|S_{I\Ib}|/\hbar$ of width $\sqrt{3n^*/\pi}$ 
(\cref{tab:sumstates}, \cref{fig:sumstates}b); as $T\to\tau_q$ the 
band moves down to $n^*\approx1$, and the sum turns into the single
exchange \eqref{eq:exchange}. Since the quartic tail reaches the 
quantum width a time of order $\tau_q$ after the core, the shift of 
the time origin in \eqref{eq:Rtime} is $O(\tau_q)$, and the quantum 
interaction carries no large factor $1/g$.

\begin{table}[t]
\centering
\small
\begin{tabular}{@{}cccccc@{}}
\toprule
$T/\tau_q$ & $S_{I\Ib}$ classical & $-\hbar\ln(\text{sum})$ 
           & $n^*$ from the sum & $\frac4\pi|S_{I\Ib}|/\hbar$ 
           & $\Delta n$ vs $\sqrt{3n^*/\pi}$\\
\midrule
0.3 & $-121.9$ & $-125.6$ & 159.0 & 155.2 & 12.9 vs 12.3\\
0.4 & $-55.1$ & $-58.2$ & 71.0 & 70.2 & 8.4 vs 8.2\\
0.6 & $-17.3$ & $-19.5$ & 22.2 & 22.0 & 4.6 vs 4.6\\
0.8 & $-7.5$ & $-9.1$ & 9.6 & 9.5 & 3.0 vs 3.0\\
\bottomrule
\end{tabular}
\caption{The $I\Ib$ interaction from the sum over the numerically
computed states of the $\sin^4(x/2)$ well (column 3), compared 
with the classical action of the stretch between the cores
(column 2). The calculation was done for $S_0/\hbar=2000$. 
The difference between columns 2 and 3 is a measure of $O(\hbar)$
corrections. The quantity $n^*$ (column 4) indicates where the 
sum over $n$ peaks, and column 5 is the saddle-point prediction  
for $n^*$. Column 6 shows the same comparison for the width of 
the distribution in $n$.}
\label{tab:sumstates}
\end{table}

In the $II$ channel the sign alternates, so each quantum carries 
the complex fugacity $-\e^{\pi}=\e^{\pi(1+i)}$. Formally the same 
saddle-point steps go through with $\pi\to\pi(1+i)$; the saddle value 
scales as the fourth power of the fugacity exponent, and $(1+i)^4=-4$ 
gives $S_{II}=+4|S_{I\Ib}|$, both the sign flip and the factor $4$ of 
\eqref{eq:interactions0}. The saddle sits below the bottom of the well, 
at the crossing orbit of positive Euclidean energy, which corresponds 
to no eigenstate. Here all states matter: the individual terms are as 
large as $\e^{+|S_{I\Ib}|/\hbar}$ while the sum is $\e^{-4|S_{I\Ib}|/
\hbar}$, and the low-lying states take part in the cancellation as much 
as those near $n^*$. The classical $II$ repulsion is an interference 
effect across the whole spectrum.

In short: for harmonic minima ``the lightest excitation sets the 
range'' and ``the classical tail sets the strength'' are the same
statement, because Poisson weights over equally spaced levels
exponentiate. For quartic minima the $I\Ib$ sum is a Boltzmann-like 
sum with fugacity $\e^{\pi}$ per quantum whose saddle point is the
classical orbit, and the $II$ sum is the same sum at the complex 
fugacity $-\e^{\pi}$.

\section{Numerical methods and checks}
\label{app:numerics}

 All numerical calculations can be done in the variable $u=x/2$, so
that $H=\frac14\big[-\frac12\partial_u^2+64G\sin^4u\big]$. This 
gives the period-$\pi$ potential $64G\sin^4u$, where the 
energies have to be divided by $4$.

\begin{itemize}
\item \emph{Bands.} We can diagonalize $H$ in the Bloch--Fourier 
basis $\e^{i(k+\theta/2\pi)x}$, in which $16G\sin^4(x/2)=6G-8G
\cos x+2G\cos2x$ is pentadiagonal. In practice we have used 
$|k|\le60$--$260$. This resolves the bandwidth up to $S_0 \approx
22$. The parameters $\chi_{\rm top}$ and $b_2$ come from a polynomial
fit of $E_0(\theta)$ in $\theta^2$ for $\theta\le0.6$. The data are 
given in \cref{tab:matching} and \cref{fig:theta}.

\item \emph{Tunneling at large $G$.} For large $G$ and large
$S_0$ we use the amplitude formula \eqref{eq:herring}. We 
numerically compute the wave function $\psi_L(\pi)$ and its
derivative. The Riccati equation $w'=2(V-E)-w^2$ for $w=
\psi'/\psi$ is integrated inwards, which is the numerically 
stable direction, from the top of the barrier (where $w=
-\sqrt{2(V(\pi)-E)}$) to $x=0$. The condition $w(0)=0$ fixes 
$E$, and $\psi=\exp\!\int w$ is normalized. The result agrees 
with the band calculation to $10^{-5}$ for $12\le S_0\le24$ 
and was used up to $S_0=2500$.

\item \emph{Quartic double well} (\cref{sec:dw}). The splittings
were computed by discretizing the Hamiltonian with fourth-order 
finite differences and diagonalizing the resulting sparse matrix. This method can
be used for $S_0\le14$. For larger $S_0$ we again use the 
amplitude formula. The wave function at the bottom of the well
is determined by Riccati integration from the top of the barrier 
and from $x\to-\infty$, matched at the minimum. The latter
method can be used for $4\le S_0\le10^6$. The two methods 
agree to $10^{-4}$ or better where they overlap. The $\ell$ 
coefficient in \eqref{eq:dwcorr} is based on $\partial_\varepsilon
\ln B_+=-0.092$. It is computed using the method explained in 
\cref{sec:corrections}, in this case applied to the perturbation 
$\varepsilon y^5$.

\item \emph{Quartic-oscillator constants.} The Riccati method 
applied to $-\frac12\partial_y^2+y^4$ gives $e_0$ and $\varphi_0$ 
out to $y=22$. The quantity $B$ is extracted from the tail. Excited
levels were determined using sparse diagonalization and finite differences.

\item \emph{Sums over states} (Appendix~\ref{app:spectral}). 
Single-well spectra were computed using a cardinal sine discrete 
variable representation (sinc DVR), and single-well states at the 
core were determined by inward Numerov integration in log-scaled 
arithmetic. Sums were computed in 260-digit decimal arithmetic
using a smooth model for $R_n$.

\item \emph{Perturbation theory around the quartic oscillator.} 
The first-order wavefunction, from the linear equation for 
$u=\varphi_1/\varphi_0$, is integrated inward from $y=14$ with 
an 8th-order method. The quantity $\partial_\varepsilon\ln B$ 
is determined from a fit in $1/y$ over $7\le y\le12$, stable to
$10^{-4}$. Finally, the second-order energy $e^{(2)}$ and $\langle y^8\rangle$ are 
computed by quadrature.

\item \emph{Fits.} The fit $t/(\sqrt2B^2G^{1/3} \e^{-S_0})
=1-0.138\,G^{-1/3}+\dots$ is extracted from the numerical 
data with the constant term reproduced to $4\times10^{-5}$.
The fit $E_0-e_0G^{1/3}=-0.04420-0.0019\,G^{-1/3}-0.0003\,
G^{-2/3}$ can be checked against the first-order value $-\tfrac16\langle 
y^6\rangle=-0.04419$.

\end{itemize}

\FloatBarrier
\bibliographystyle{unsrtnat-eprint}
\bibliography{refs}

\end{document}